\documentstyle[aps,prl,epsfig,psfig]{revtex}

\newcommand{\be}{\begin{equation}}
\newcommand{\ee}{\end{equation}}
\newcommand{\bea}{\begin{eqnarray}}
\newcommand{\eea}{\end{eqnarray}}
   \newcommand{\upar}{\uparrow}
   \newcommand{\dn}{\downarrow}
   \newcommand{\ek}{\varepsilon_k}
   \newcommand{\ef}{\varepsilon_f}

\begin{document}
\widetext
  \leftskip 10.8pt
  \rightskip 10.8pt

\title{The Kondo Lattice Model}
\author{M. Lavagna$^{1,*}$ and C. P\'epin$^2$}
\address{$^1$Commissariat \`a l'Energie Atomique, DRFMC /SPSMS, 
17, rue des Martyrs,
   38054 Grenoble Cedex 9, France}
\address{$^2$University of Oxford, Department of Physics, 1 Kable Road, OX1 3NP Oxford, UK}

\date{\today} 
\maketitle \widetext
  \leftskip 10.8pt
  \rightskip 10.8pt
  \begin{abstract}
In this lecture, we review the experimental situation of heavy Fermions with emphasis on the existence of a quantum phase transition (QPT) and related non-Fermi liquid (NFL) effects. We overview the Kondo lattice model (KLM) which is believed to describe the physics of those systems. After recalling the existing theories based on large-N expansion and various N=2 schemes, we present two alternative approaches: (i) a spin fluctuation-Kondo functional integral approach treating the spin-fluctuation and Kondo effects on an equal footing, and (ii) a supersymmetric theory enlarging the usual fermionic representation of the spin into a mixed fermionic-bosonic representation in order to describe the spin degrees of freedom as well as the Fermi-liquid type excitations. This kind of approaches may open up new prospects for the description of the critical phenomena associated to the quantum phase transition in Heavy-Fermion systems. 
\end{abstract}

\bigskip

Lecture given at the XXXVIII Cracow School of Theoretical Physics, Zakopane, June 1998: "New Quantum Phases, Elementary Excitations and Renormalization in High Energy and Condensed Matter Physics" (to appear in Acta Physica Polonica B 1999).

\bigskip

I. Experimental overview
\par
II. Theoretical overview
\par
\hangindent=1cm
\hangafter=-1
A. The large-N expansion
\par
\hangindent=1cm
\hangafter=-1
B. The different N=2 approaches
\par
III. The spin fluctuation-Kondo functional integral approach
\par
\hangindent=1cm
\hangafter=-1
A. Saddle-point
\par
\hangindent=1cm
\hangafter=-1
B. Gaussian fluctuations
\par
\hangindent=1cm
\hangafter=-1
C. Dynamical spin susceptibility
\par
\hangindent=1cm
\hangafter=-1
D. Physical discussion
\par
IV. The supersymmetric approach

\section{\protect\bigskip Experimental overview}
In heavy-Fermion systems with 4f or 5f atoms (such as Ce or U), the
proximity of the electronic orbital to the Fermi level confers a Kondo
effect at low temperature, i.e. an on-site compensation of localized
magnetic moment by conduction electrons \cite{lee}. A direct consequence is the
observation at low temperature of a very large effective electronic mass m*
derived from the huge linear specific heat coefficient $\gamma =C/T$ and a
correspondingly large Pauli susceptibility. Simultaneously, the realization
of the de Haas-van Alphen (dH-vA) quantum oscillations \cite{julian92} also concludes in
favour of the existence of heavy quasiparticles.

In addition to the Kondo effect, those systems are characterized by
long-range RKKY (Ruderman-Kittel-Kasuya-Yoshida) interactions between
neighboring local moments mediated by conduction electrons. The competition
between the Kondo effect and the RKKY interactions leads to the possibility
of either a non-magnetic or a long-range magnetically-ordered ground state \cite{doniach}.
A zero-temperature quantum phase transition occurs governed by the value of
the exchange coupling J between the spin of the conduction electron and the
local moment. One of the most striking properties of the heavy-Fermion
compounds discovered these last years is the experimental possiblity to
explore this quantum phase transition \cite{lohneysen,julian96,steglich,kambe} by varying the composition change (as
in $CeCu_{6-x}Au_{x}$ or $Ce_{x}La_{1-x}Ru_{2}Si_{2}$), or by applying a
pressure or a magnetic field. Thus a magnetic instability is observed at $%
x_{c}=0.1$ in $CeCu_{6-x}Au_{x}$ \cite{lohneysen}, and $x_{c}=0.08$ in $%
Ce_{x}La_{1-x}Ru_{2}Si_{2}$ \cite{kambe}. For $x=x_{c}$ where $T_{N}=0$, the observed
behavior at low temperature is at odds with that usually expected for a
simple Fermi liquid (FL). In $CeCu_{6-x}Au_{x}$, the specific heat $C$
depends on T as $C/T\sim -Ln(T/T_{0})$, the magnetic susceptibility as $\chi
\sim 1-\alpha \sqrt{T}$, and the T-dependent part of the resistivity as $%
\Delta \rho \sim T$ (instead of $C/T\sim \chi \sim Const$ and $\Delta \rho
\sim T^{2}$ in the Fermi liquid state). Pressure or large magnetic fields
are found to restore the FL behavior. 

The origin of this non-Fermi liquid
(NFL) regime is a largely discussed problem. The three main interpretations
which have been proposed rely on $(i)$ a single impurity multichannel Kondo
effect in which the internal degree of freedom is provided by the 4f or 5f
quadrupolar moment \cite{cox}, $(ii)$ a distribution of the Kondo coupling due to the
disorder leading to a distribution of the Kondo temperature $P(T_{K})$ \cite{miranda} and $%
(iii)$ the proximity of a quantum phase transition \cite{hertz,millis93,moriya,continentino,rosch,pepin} as is emphasized in this
course. 

Another important insight is provided by the Inelastic Neutron
Scattering (INS) experiments carried out in systems close to the magnetic
instability. The measurements performed in pure compounds $CeCu_{6}$ or $%
CeRu_{2}Si_{2}$ \cite{regnault,aeppli} have shown the presence of two distinct contributions to the
dynamic magnetic structure factor: a q-independent quasielastic component,
and a strongly q-dependent inelastic contribution peaked at the value $%
\omega _{\max }$ of the frequency. The same experiments carried out in
systems with varying concentrations as $Ce_{x}La_{1-x}Ru_{2}Si_{2}$ \cite{raymond} , show a
shift of $\omega _{\max }$ to zero when getting near the magnetic
instability. Any theory aimed to describe the quantum critical phenomena in
heavy-Fermion compounds should account for the so-quoted behavior of the
dynamical spin susceptibility.

\bigskip

\section{\protect\bigskip Theoretical overview}

The model which is believed to describe the heavy-Fermion systems is the
Periodic Anderson Model (PAM) defined for the case of spin $1/2$ \cite{anderson} as
\begin{equation}
\label{pam}
H=\sum_{k\sigma}\varepsilon _{k}c_{k\sigma }^{\dagger}c_{k\sigma}+E_{0}\sum_{i\sigma}f_{i\sigma }^{\dagger}f_{i\sigma }+V%
\sum_{i\sigma}(c_{i\sigma }^{\dagger}f_{i\sigma }+c.c.)+U%
\sum_{i} n_{fi}^{\upar}\,n_{fi}^{\dn}
\end{equation}

where $n_{fi}^{\sigma }=f_{i\sigma }^{+}f_{i\sigma }$. It describes the
conduction electrons $c_{k\sigma }$ with dispersion $\varepsilon _{k}$ which
hybridize with the localized $f$ electrons of energy $E_{0}$. The grand
canonical ensemble is used and both energies $\varepsilon _{k}$ and $E_{0}$
are measured from the chemical potential $\mu$. The hybridization matrix
element is approximated by a constant $V$. $U$ represents the on-site
Coulomb repulsion between $f$ electrons.

In the Kondo limit, a canonical transformation allows to change the Periodic
Anderson Model into the Kondo lattice model (KLM) defined as
\begin{equation}
\label{klm}
H=\sum_{k\sigma}\varepsilon _{k}c_{k\sigma }^{\dagger}c_{k\sigma
}+J\sum_{i}{\bf{s}}_{i}.{\bf{S}}_{i}
\end{equation}

where $J$ is the on-site Kondo coupling between the spin of the conduction
electrons ${\bf{s}}_{i}=\sum\limits_{\sigma \sigma
^{\prime }}c_{i\sigma }^{\dagger}\mbox{\boldmath{$\tau$}}_{\sigma \sigma ^{\prime
}}c_{i\sigma ^{\prime }}$ and the localized spin represented by ${\bf{S}}_{i}=\sum\limits_{\sigma \sigma
^{\prime }}f_{i\sigma }^{\dagger}\mbox{\boldmath{$\tau$}}_{\sigma \sigma ^{\prime
}}f_{i\sigma ^{\prime }}$ in the
Abrikosov pseudo-fermionic representation of the spin imposing the
constraint $n_{fi}=\sum_{\sigma }n_{fi}^{\sigma }=1$. That
canonical transformation analogous to the Schrieffer-Wolf transformation for
the single-impurity case is valid in the regime $|\varepsilon
_{k}|<<(-E_{0}) $ and $|\varepsilon _{k}|<<|E_{0}+U|$. One gets: $J=V^{2}%
\left[ -\frac{1}{E_{0}}+\frac{1}{E_{0}+U}\right] $. In the $U\rightarrow
\infty $ limit, one has $J=-V^{2}/E_{0}$ while in the symmetric Anderson
model defined by $E_{0}=-U/2$, the result is $J=4V^{2}/U$. The Kondo lattice 
Model has been first introduced by Doniach in 1977 \cite{doniach} and we refer to the paper
of Tsunetsugu et al \cite{tsunetsugu} for an extensive review on the KLM essentially at D=1.

Let us first recall 
the main physical ideas behind the Kondo Lattice model: $%
(i)$ the competition between the Kondo effect and the RKKY interactions
leading to the so-called ''Doniach phase diagram'' and $(ii)$ the nature of
the screening of the local moments in the lattice. 

Concerning the point $(i)$%
, the competition between the Kondo effect on each site which tends to
suppress the magnetic moment with decreasing temperature and the RKKY
interactions which, on the contrary, tend to magnetically order the local
moments, leads to the well-known Doniach phase diagram \cite{doniach}. Let us call $T_{K}^{0}$ the Kondo temperature for the
single-impurity and $T_{N}^{0}$ (or $T_{C}^{0}$) the N\'{e}el (or Curie)
temperatures in the absence of Kondo effect: $T_{K}^{0}=D\exp \left( -1/\rho
_{0}J\right) $ and $T_{N}^{0}$ $\sim \left( \rho _{0}J\right) ^{2}$ where $D$
and $\rho _{0}$ are the bandwidth and the density of states at the Fermi
level of the conduction electron band. Thus, at small $\rho _{0}J$, $%
T_{N}^{0}$ is larger than $T_{K}^{0}$ and a long-range magnetic order is
established with eventually a reduction of the magnetic moment due to the
Kondo effect. Oppositely, at large $\rho _{0}J$, $T_{K}^{0}$ is larger than $%
T_{N}^{0}$, the Kondo effect wins and the system does not order
magnetically. Therefore, the real N\'{e}el temperature $T_{N\text{ }}$ first
increases with increasing $\rho _{0}J$, passes through a maximum and finally
goes to zero at a critical value of the coupling $\rho _{0}J_{C}$ giving
rise to a zero-temperature quantum critical point. 

As far as the nature of
the screening is concerned (point $(ii)$), an important idea advanced by
Nozi\`{e}res \cite{nozieres} is the possibility of an exhaustion of the conduction electrons
in the screening of the local moments. In the single-impurity Kondo case, at
low temperature, the local moment is screened by the conduction electrons
and a spin-singlet state is formed. In this so-called Kondo effect, the
number of conduction electrons in the screening cloud formed around the
impurity is equal to 1. In the Kondo lattice case, the number of conduction
electrons which are available are to be taken within a thermal window of
width $k_{B}T$ around the Fermi level. It should be compared to the number
of sites $N_{S}$ to be screened. Depending on the value of the parameters,
some situations may occur where the available conduction electrons are
''exhausted'' before achieving complete screening leaving residual
unscreened spin degrees of freedom on the impurities. That idea of 
''uncomplete'' Kondo effect is at the root of the
supersymmetric theory that we propose later on.

At high temperature, pertubation techniques in J may be applied leading to
the famous Kondo minimum in the resistivity as a function of temperature.
Those pertubation techniques fail below the Kondo temperature and there is a
need for other techniques to solve the problem at low-temperature. The other
approaches based on the Bethe Ansatz and the Renormalizaion Group which
revealed very powerful in the single-impurity case cannot be generalized to
the lattice case. In that context, there has been an intensive search for
new approaches among which the large-N and more generally the functional
integral approaches described here.

\bigskip

\subsection{\protect\bigskip The Large-N expansion}

An important breakthrough in the understanding of the periodic Anderson
Model occured about 15 years ago when the idea of slave-bosons was
introduced: \cite{coleman,read83} for the single-impurity case, 
\cite{millis87,auerbach86,lavagna87,read84} for the
lattice. In the limit of large on-site Coulomb repulsion ($U\rightarrow
\infty )$ where the double-occupancy is energetically forbidden, one can
introduce a slave-boson representation in which the two allowed states i.e.
the empty or the singly-occupied states are represented by $e_{i}^{\dagger}|0>$
and $f_{i\sigma }^{\dagger}|0>$. The exclusion of double-occupancy is expressed as

\begin{equation}
\label{constr}
P_{i}=e_{i}^{\dagger}e_{i}+\sum_{\sigma }f_{i\sigma }^{\dagger}f_{i\sigma }-1=0
\end{equation}

The physical electron creation operator which creates transitions between
empty and singly-occupied sites is represented by $f_{i\sigma }^{\dagger}e_{i}$,
while the number of $f$ electrons on site i of spin $\sigma $ is equal to $%
f_{i\sigma }^{\dagger}e_{i}e_{i}^{\dagger}f_{i\sigma }=f_{i\sigma }^{\dagger}f_{i\sigma }$
provided that the local constraint is satisfied. Hence the slave-boson
representation of the $U\rightarrow \infty $\ PAM hamiltonian is given by
\begin{equation}
\label{slave}
H=\sum_{k,\sigma }\varepsilon _{k}c_{k\sigma }^{\dagger}c_{k\sigma
}+E_{0}\sum_{i,\sigma }f_{i\sigma }^{\dagger}f_{i\sigma }+V%
\sum_{i,\sigma }(c_{i\sigma }^{\dagger}e_{i}^{\dagger}f_{i\sigma }+c.c.)
\end{equation}

provided that the local constraint Eq.(\ref{constr}) is satisfied that is enforced with the
aid of a time-independent Lagrange multiplier $\lambda _{i}$. The operators $e_{i}^{+}$ and
$f_{i\sigma }^{+}$ obey bosonic and fermionic statistics respectively. 

It is then convenient to generalize the original PAM model from $SU(2)$ to $%
SU(N)$ by allowing for the spin index $\sigma $ in Eq.(\ref{slave}) to run from $-S$ to $S
$. That corresponds to the situation of impurities of spin $S$ coupled to
conduction electrons of degeneracy $N$ with $N=2S+1\ $. The corresponding $SU(N)$ model is
interesting because it can be solved exactly in the limit $N\rightarrow
\infty $. For this limit to make sense, the hybridization matrix element $V$
should be scaled as $1/\sqrt{N}$ therefore $V=\widetilde{V}/\sqrt{N}$. 

We do
not give here all the details of the calculations which can be found in the
litterature. Let us say that the saddle-point approximation which
consists to take $e_{i}$ and $\lambda _{i}$ as site-independent (as well as
time-independent for $e_{i}$) is exact in the limit $N\rightarrow \infty $.It leads to the formation of two quasiparticle bands of energies $E_{k}^{\pm
}=\frac{1}{2}\varepsilon _{k}+\varepsilon _{f}\pm \sqrt{\left( \varepsilon
_{k}-\varepsilon _{f}\right) ^{2}+4V^{2}e_{0}^{2}}$ where $\varepsilon
_{f}=E_{0}+\lambda _{0}$. The values of $e_{0},$ $\varepsilon _{f}$ and $\mu 
$ are fixed by the saddle-point equations. The two bands are split by a
hybridization gap and the density of states at the Fermi level is strongly
renormalized $\rho (E_{F})\sim 1/T_{K}$ where $T_{K}=D\exp (E_{0}/\rho
_{0}V^{2})$. The specific heat coefficient $\gamma $ and the magnetic
susceptibility $\chi $ are also strongly enhanced with a Wilson ratio $\chi
/\gamma $ equal to 1. Oppositely, the charge susceptibility is found to be
unenhanced. The ground state corresponds to a collective Kondo screening in
which only a fraction of conduction electons equal to $T_{K}/D$ screens each
of the impurities. 

Next step is to include the gaussian fluctuations around
the saddle-point. The corresponding corrections in $1/N$ generate effective
interactions among the quasiparticles which can be analyzed in terms of
Landau paramaters. In the case of the multichannel single-Kondo impurity
problem, non-crossing approximation methods \cite{kroha} have been extensively used to
derive the finite temperature behavior with very accurate comparison with
the Bethe-Ansatz and Conformal Field theory results. The generalization to
the case of the lattice has still to be done. All the results can be
reproduced by starting instead from the $N\rightarrow \infty $ Kondo lattice
hamiltonian and performing a Hubbard Stratonovich transformation on the
coupling term making the field $\Phi $ appear. There is then a one-to-one
equivalence between $\Phi _{0}=J\sum_{k\sigma }<c_{k\sigma }^{\dagger}f_{k\sigma }>
$ and $Ve_{0}$. In all cases, no magnetic instability is found at the order $%
1/N$ since the RKKY\ interactions occur at the order $1/N^{2}$ \cite{houghton}. The latter
point constitutes a serious drawback of the large-$N$ expansion which makes
it inappropriate to describe the Quantum Critical Point observed
experimentally. The fact that the slave-boson does not carry spin implies
that spin and charge fluctuations cannot be treated on an equal footing
contrary to what happens with other slave-boson representations as the one
introduced by Kotliar and Ruckenstein that we present now.

\bigskip

\subsection{\protect\bigskip The different N=2 approaches}
In the case of a degeneracy $N=2$ ($S=1/2$), Kotliar and Ruckenstein (KR) \cite{kotliar}
introduced 4 slave-bosons $e_{i}$, $p_{i\sigma }$ and $d_{i}$ in order to
keep track of the 4 possible local configurations so that the empty $|O>_{i}$%
, the singly-occupied $|\sigma >_{i}$ and the doubly-occupied $|\uparrow
\downarrow >_{i}$ states are represented by: $|O>_{i}=e_{i}^{\dagger}|vac>$, $%
|\sigma >_{i}=p_{i\sigma }^{\dagger}f_{i\sigma }^{\dagger}|vac>$ and $|\uparrow
\downarrow >_{i}=d_{i}^{\dagger}f_{i\uparrow }^{\dagger}f_{i\downarrow }^{\dagger}|vac>$ where 
$e_{i}$, $p_{i\sigma }$ and $d_{i}$ are bosons and $f_{i\sigma }$
fermions. Then the PAM can be written as

\begin{equation}
H=\sum_{k,\sigma }\varepsilon _{k}c_{k\sigma }^{\dagger}c_{k\sigma
}+E_{0}\sum_{i,\sigma }f_{i\sigma }^{\dagger}f_{i\sigma }+V%
\sum_{i,\sigma }(c_{i\sigma }^{\dagger}f_{i\sigma }z_{i\sigma
}+c.c.)+U\sum_{i}d_{i}^{\dagger}d_{i}
\end{equation}

with \[z_{i\sigma }=\left( e_{i}^{\dagger}p_{i\sigma }+p_{i-\sigma }^{\dagger}d_{i}\right) /%
\left[ \sqrt{1-e_{i}^{\dagger}e_{i}-p_{i-\sigma }^{\dagger}p_{i-\sigma }}\sqrt{%
1-d_{i}^{\dagger}d_{i}-p_{i\sigma }^{\dagger}p_{i\sigma }}\right] \]

provided that the 3 following constraints are fulfilled

\begin{equation}
P_{i}=e_{i}^{\dagger}e_{i}+\sum_{\sigma }p_{i\sigma }^{\dagger}p_{i\sigma }
-1=0
\end{equation}

\[Q_{i\sigma }=f_{i\sigma }^{\dagger}f_{i\sigma }-(p_{i\sigma }^{\dagger}p_{i\sigma
}+d_{i}^{\dagger}d_{i})=0\]

The choice of the denominator in $z_{i\sigma }$ guarantees to recover the
free electron gaz limit at $U\rightarrow 0$. This representation first
introduced in the case of the Hubbard model was shown at the saddle-point
level to give back the
variational Gutzwiller approximation (GA) as developed by Rice and Ueda \cite{rice}
. It then allows to include the gaussian fluctuations around the GA
solution \cite{li}, \cite{lavagna90}. In the case of the PAM
\cite{yang,sun,doradzinski,moller}, the KR representation already leads to
interesting results at the saddle-point level as soon as staggered
symmetry-broken state appropriate for bipartite lattice with nesting is
allowed. Notably, at d=1 \cite{yang}, the approach essentially gives the same results as
those obtained by the variational wave function approach of Gulacsi, Strack
and Vollhardt \cite{gulacsi}. At infinite dimension \cite{sun}, a phase transition to the
antiferromagnetic insulator is found below a critical value $V_{C}$ of the
hybridization matrix element consistent with the Doniach predictions and in
quantitative agreement with the $d=\infty $ QMC \cite{jarrell} and exact diagonalization
\cite{rozenberg} results. The general phase diagram for the three-dimensional case has been
determined by Doradzinski and Spalek \cite{doradzinski}. Those studies enlighten on the nature
of the moment compensation which takes place in the antiferromagnetic state.
The study of the V-dependence of the staggered magnetizations $m_{f}$ and $%
m_{c}$\ shows an almost total moment compensation of $m_{f}$ and $m_{c}$
near $V_{C}$ suggesting an itinerant magnetism in which f and c electrons
are part of the same quasiparticles. Oppositely, in the $V\rightarrow 0$
limit, $m_{f}$ saturates while $m_{c}$ goes to zero indicating a local
moment magnetism. The figure 4 of the paper \cite{doradzinski} illustrates 
the latter point by showing the
V-dependence of $m_{f}$ and $m_{c}$ in the antiferromagnetic insulating
state.

Finally, other path integral approaches to the KLM have been proposed based
on different Hubbard Stratonovich decouplings of the exchange term \cite{lacroix,andrei}. For the
extended KLM\ in which the Heisenberg interactions among neighboring sites
are included, we will mention the work of Coleman and Andrei \cite{andrei} which consists
to keep the Resonant-Valence-Bond (RVB) parameter $\chi _{ij}=J\sum_{\sigma
}f_{i\sigma }^{\dagger}f_{j\sigma }$ on neighboring sites at the same time as the
Kondo parameter $\Phi =J\sum_{\sigma }c_{i\sigma }^{\dagger}f_{i\sigma }$ quoted
before. This approach leads to the stabilization of a spin-liquid state at
low temperature with possible anisotropic superconducting instability. This
approach has been used in a recent paper by Iglesias, Lacroix and Coqblin \cite{iglesias}
where they propose a revisited version of the Doniach phase diagram in which
the Kondo temperature is drastically reduced resulting of the formation of
the resonant valence bonds.

In the rest of the paper, we will develop two alternative approaches to the
Kondo lattice model: (i) the first one consists to keep the f and c
magnetizations at the same level as the Kondo parameter \cite{pepin}. We will show how
it is possible to account for the spin-fluctuation and the Kondo effects on
an equal footing thus combining both large N and spin-fluctuation theories. To
our point of view, this approach constitutes an ideal framework to study the
quantum critical phenomena around the magnetic transition; (ii) the second
approach consists in enlarging the usual Abrikosov pseudo-fermionic
representation of the spin into a mixed fermionic-bosonic representation in
order to describe the spin degrees of freedom as well as the Fermi-liquid
type excitations \cite{lavagna97}. The analogy of the approach with the supersymmetry theory
of disordered systems leads to give it the nickname of ''supersymmetric
approach''. 

\bigskip

\section{\protect\bigskip The spin fluctuation-Kondo functional Integral Approach}

In the grand canonical ensemble, the hamiltonian of the Kondo lattice 
model (KLM) constituted by a periodic array of
Kondo impurities with an average number of conduction electrons per site
$n_c$ is
written as

\begin{equation}
\label{eq1}
H=\sum_{k\sigma }\varepsilon _{k}c_{k\sigma }^{\dagger}c_{k\sigma
}+J\sum\limits_{i}{\bf{S}}_{i} \cdot \sum\limits_{\sigma \sigma ^{\prime
}}c_{i\sigma }^{\dagger}\mbox{\boldmath{$\tau$}}_{\sigma \sigma'} c_{i\sigma ^{\prime } }-\mu N_S(\frac{1}{N_{S}}%
\sum_{k\sigma }c_{k\sigma }^{\dagger}c_{k\sigma }-n_{c})
\end{equation}

in which $\mbox{\boldmath{$\tau$}}$ are the Pauli matrices $%
\left(\mbox{\boldmath{$\tau$}} ^{x}, \mbox{\boldmath{$\tau$}} ^{y}, 
\mbox{\boldmath{$\tau$}} ^{z}\right) $ and $\mbox{\boldmath{$\tau$}}^{0}$ the unit matrix; J is the antiferromagnetic
Kondo interaction $\left( J>0\right) $.

We use the Abrikosov pseudo-fermion representation of the spin $%
{\bf{S}}_{i}$: ${\bf{S}}_{i}=\sum\limits_{\sigma \sigma
^{\prime }}f_{i\sigma }^{\dagger}\mbox{\boldmath{$\tau$}}_{\sigma \sigma ^{\prime
}}f_{i\sigma ^{\prime }}$. The projection into the physical subspace is
implemented by a local constraint  

\be
\label{eq2}
Q_{i}=\frac{1}{N_{S}}%
\sum\limits_{i\sigma }f_{i\sigma }^{+}f_{i\sigma }-1=0
\ee

A Lagrange multiplier $\lambda _{i}$ is introduced to enforce the local
constraint $Q_{i}$. Since $[Q_{i},H]=0$, $\lambda _{i}$ is time-independent.

In this representation, the partition function of the KLM can be expressed
as a functional integral over the coherent states of the fermion fields 

\begin{equation}
\label{eq3}
Z=\int {\cal D}c_{i\sigma }{\cal D}f_{i\sigma }d\lambda _{i}\exp \left[-\int_{0}^{\beta }{\cal L}(\tau
)d\tau\right]  
\end{equation}

where the Lagrangian ${\cal L}(\tau)$ is given by

\[
{\cal L}(\tau )={\cal L}_{0}(\tau )+H_{0}(\tau )+H_{J}(\tau ) 
\]

\[
{\cal L}_{0}(\tau )=\sum_{i\sigma }c_{i\sigma }^{\dagger}\partial _{\tau }c_{i\sigma
}+f_{i\sigma }^{\dagger}\partial _{\tau }f_{i\sigma }
\]

\[
H_{0}(\tau )=\sum_{k\sigma }\epsilon _{k}c_{k\sigma }^{\dagger}c_{k\sigma }-\mu
N_{S}\left( \frac{1}{N_{S}}\sum_{k\sigma }c_{k\sigma }^{\dagger}c_{k\sigma
}-n_{c}\right) +\sum_{i}\lambda _{i}Q_i 
\]

\[
H_{J}(\tau )=J\sum_{i}{\bf{S}}_{fi} \cdot {\bf{S}}_{ci}
\]
with ${\bf{S}}_{c_{i}}=\sum\limits_{\sigma \sigma ^{\prime
}}c_{i\sigma }^{\dagger}\mbox{\boldmath{$\tau$}}_{\sigma \sigma ^{\prime
}}c_{i\sigma ^{\prime }}$ and ${\bf{S}}_{fi}={\bf{S}}_{i}$

\par
We perform a Hubbard-Stratonovich transformation on the Kondo
interaction term $H_{J}(\tau )$. Since more than one field is implied in the transformation, 
an uncertainty is left on the way of decoupling. We propose to remove it in the following way. 
First, we note that $H_{J}(\tau)$ may also be written as 

\be
\label{eq4}
H_{J}(\tau )=-\frac{3J}{8}\sum_{i}n_{fc_{i}}n_{cf_{i}}+\frac{J}{2}\sum_{i}%
{\bf{S}}_{fc_{i}} \cdot {\bf{S}}_{cf_{i}}
\ee

where ${\bf{S}}_{fc_{i}}=\sum\limits_{\sigma \sigma ^{\prime
}}f_{i\sigma }^{\dagger}\mbox{\boldmath{$\tau$}}_{\sigma \sigma ^{\prime
}}c_{i\sigma ^{\prime }}$ and $n_{fc_{i}}=\sum\limits_{\sigma \sigma ^{\prime }}f_{i\sigma }^{\dagger} {\tau}_{\sigma \sigma ^{\prime}}^{0} c_{i\sigma ^{\prime}}$
(respectively ${\bf{S}}_{cf_{i}}$ and $n_{cf_{i}}$ their hermitian conjugate).

\bigskip 
The Kondo interaction term is then given by any linear combination of 
$J\sum\limits_{i}{\bf{S}}_{fi}\cdot {\bf{S}}_{ci}$ (with a
weighting factor x) and of the term appearing in the right-hand side of
Equation (\ref{eq4}) (with a weighting factor (1-x)). x is chosen so as to recover the
usual results obtained within the slave-boson theories \cite{millis87,auerbach86,lavagna87,read84}. 
One can check that this is
the case for $x=1/3$.
The Kondo interaction term is then given by 

\be
\label{eq5}
H_{J}(\tau )=J_{S}\sum_{i}\left( {\bf{S}}_{f_{i}}\cdot {\bf{%
S}}_{c_{i}}+{\bf{S}}_{fc_{i}}\cdot {\bf{S}}%
_{cf_{i}}\right) -J_{C}\sum_{i}n_{fc_{i}}n_{cf_{i}}
\label{cinq}
\ee

with $J_{S}=J/4$ and $J_{C}=J/3$.

\bigskip
Performing a generalized Hubbard-Stratonovich transformation on the
partition function Z makes the fields $\Phi _{i}$, $\Phi
_{i}^{\ast }$ (for charge) and $\mbox{\boldmath{$\xi$}}_{f_{i}},$ $%
\mbox{\boldmath{$\xi$}}_{c_{i}}$ appear (omitting the fields associated to $%
{\bf{S}}_{fc_{i}},$ ${\bf{S}}_{cf_{i}})$. We get

\be
\label{eq6}
Z=\int d\Phi _{i}d\Phi _{i}^{\ast }d\mbox{\boldmath{$\xi$}}_{f_{i}}d%
\mbox{\boldmath{$\xi$}}_{c_{i}}{\cal D}c_{i\sigma }{\cal D}f_{i\sigma }d\lambda _{i}\exp \left[-\int_{0}^{\beta }{\cal L}^{\prime}(\tau
)d\tau\right]  
\ee
with 
\[
{\cal L}^{\prime}(\tau )={\cal L}_{0}(\tau )+H_{0}(\tau )+H^{\prime}_{J}(\tau ) 
\]

\[
H^{\prime }_{J}(\tau )=\sum_{i\sigma \sigma ^{\prime }}
\left( 
\begin{array}{cc}
c_{i\sigma }^{\dagger}, & f_{i\sigma }^{\dagger}
\end{array}
\right) 
\left( 
\begin{array}{cc}
-J_{S}i\mbox{\boldmath{$\xi$}}_{f_{i}}\cdot \mbox{\boldmath{$\tau$}}_{\sigma \sigma ^{\prime
}} & J_{C}\Phi
_{i}^{\ast}\tau^{0}_{\sigma \sigma ^{\prime
}} \\ 
J_{C}\Phi _{i}\tau^{0}_{\sigma \sigma ^{\prime
}} & -J_{S}i\mbox{\boldmath{$\xi$}}_{c_{i}}\cdot 
\mbox{\boldmath{$\tau$}}_{\sigma \sigma ^{\prime
}}
\end{array}
\right) 
\left( 
\begin{array}{c}
c_{i\sigma ^{\prime }} \\ 
f_{i\sigma ^{\prime }}
\end{array}
\right)  +J_{C}\sum_{i}\Phi _{i}^{\ast }\Phi _{i}+J_{S}\sum_{i}%
\mbox{\boldmath{$\xi$}}_{f_{i}}.\mbox{\boldmath{$\xi$}}_{c_{i}}
\]

\subsection{\protect\bigskip Saddle-Point}

The saddle-point solution is obtained for space and time independent fields $%
\Phi _{0\text{ }}$, $\lambda _{0}$, $\xi _{f_{0}}$ and $\xi _{c_{0}}$. 
In the magnetically-disordered regime 
($\xi _{f_{0}}=\xi _{c_{0}}=0)$, it leads to renormalized bands $\alpha 
$ and $\beta $ as schematized in Figure 1. Noting  $\sigma _{0}^{(\ast)}=J_C \Phi _{0}^{(\ast)}$ and
$\varepsilon _{f}=\lambda _{0}$, $\alpha _{k\sigma }^{\dagger}|0>$ and $%
\beta _{k\sigma }^{\dagger}|0>$ are the eigenstates of

\begin{equation}
\label{eq7}
{\bf G}_{0}^{-1 \sigma }(\bf{k},\tau )=\left( 
\begin{array}{cc}
\partial _{\tau }+\varepsilon _{k} & \sigma _{0}^{\ast} \\ 
\sigma _{0} & \partial _{\tau }+\varepsilon _{f}
\end{array}
\right) 
\end{equation}
with respectively the eigenenergies $\left( \partial _{\tau }+E_{k}^{-}\right)$ 
and $\left( \partial _{\tau }+E_{k}^{+}\right) $. In the notations: 
$x_{k}=\varepsilon _{k}-\varepsilon _{f}$, 
$y_{k}^{\pm }=E_{k}^{\pm }-\varepsilon _{f}$ and $\Delta _{k}=\sqrt{x_{k}^{2}+4\sigma _{0}^{2}}$, we get

\begin{equation}
y_{k}^{\pm }=\left( x_{k}\pm \Delta _{k}\right) /2
\end{equation}
Let us note $U_{k\sigma }^{\dagger}$ the matrix transforming the initial basis $%
(c_{k\sigma }^{\dagger},$ $f_{k\sigma }^{\dagger})$ to the eigenbasis $(\alpha _{k\sigma }^{\dagger},$ $%
\beta _{k\sigma }^{\dagger})$. The hamiltonian being hermitian, the matrix $U_{k\sigma
}$ is unitary : $U_{k\sigma }U_{k\sigma }^{\dagger}=U_{k\sigma }^{\dagger}U_{k\sigma }=1$. In
the notation $U_{k\sigma }^{\dagger}=\left( 
\begin{array}{cc}
-v_{k} & u_{k} \\ 
u_{k} & v_{k}
\end{array}
\right) $, we have 
\be
\label{eq9}
u_{k}=\frac{-\sigma _{0}/y_{k}^{-}}{\sqrt{1+(\sigma _{0}/y_{k}^{-})^{2}}}=%
\frac{1}{2}\left[ 1+\frac{x_{k}}{\Delta _{k}}\right] 
\ee
\[
v_{k}=\frac{1}{\sqrt{1+(\sigma _{0}/y_{k}^{-})^{2}}}=\frac{1}{2}\left[ 1-%
\frac{x_{k}}{\Delta _{k}}\right] 
\]
The saddle-point equations together with the conservation of the number of
conduction electrons are written as 
\begin{equation}
\label{eq10}
\sigma _{0}=\frac{1}{N_{S}}J_{C}\sum_{k\sigma }u_{k}v_{k}~n_{F}(E_{k}^{-})
\end{equation}
\[
1=\frac{1}{N_{S}}\sum_{k\sigma }u_{k}^{2}~n_{F}(E_{k}^{-}) 
\]
\[
n_{c}=\frac{1}{N_{S}}\sum_{k\sigma }v_{k}^{2}~n_{F}(E_{k}^{-}) 
\]
Their resolution leads to
\be
\label{eq11}
\left| y_{F}\right| =D\exp \left[ -2/\left( \rho _{0}J_C\right) \right] 
\ee
\[
2\rho _{0}\sigma _{0}^{2}/\left| y_{F}\right| =1 
\]
\[
\mu =0 
\]
where $y_{F}=\mu -\varepsilon _{F}$ and $\rho _{0}$ is the bare density of
states of conduction electrons ($\rho _{0}=1/2D$ for a flat band). Noting 
$y=E-\varepsilon _{F}$, the density of states at the energy E is $\rho \left( E\right) =\rho _{0}\left( 1+\sigma _{0}^{2}/y^{2}\right)$. If $n_c<1$, the chemical potential is located just below the upper edge 
of the $\alpha$-band. The system is metallic. The density of states at the Fermi level 
is strongly enhanced towards the bare density of states of conduction electrons : 
$\rho (E_{F})/\rho_{0}=(1+\sigma _{0}^{2}/y_{F}^{2})\sim 1/(2\rho _{0}\left| y_{F}\right| )$.
That corresponds to the flat part of the $\alpha$-band in Figure 1. It is
associated to the formation of a Kondo or Abrikosov-Suhl resonance pinned at the Fermi level resulting of the Kondo effect. The low-lying excitations are 
quasiparticles of large effective mass $m^{\ast }$ as observed in heavy-Fermion
systems. Also note the presence of a
hybridization gap between the $\alpha $ and the $\beta $ bands. The direct
gap of value $2\sigma _{0}$ is much larger than the indirect gap equal to 2$%
\left| y_{F}\right| $. The saddle-point solution transposes to N=2 the
large-N results obtained within the slave-boson mean-field theories \cite{millis87,auerbach86,lavagna87,read84}.

\begin{figure}
\centerline{\psfig{file=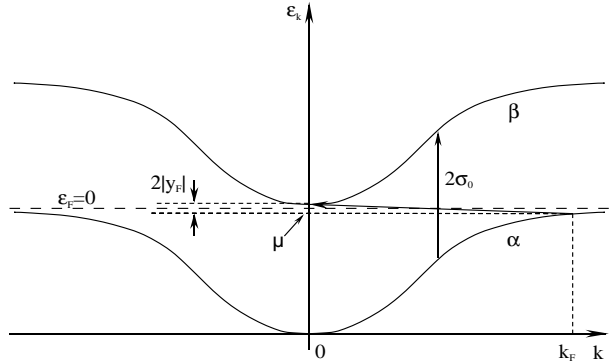,height=5cm,width=8cm}}
\caption{Energy versus wave-vector k for the two bands $\alpha$ and $\beta$. Note 
the presence of a direct gap of value $2\sigma _{0}$ and of an indirect gap of
value $2\left| y_{F}\right|$.}
\label{fig1}
\end{figure}

\subsection{\protect\bigskip Gaussian fluctuations}

We now consider the gaussian fluctuations around the saddle-point solution.
Following Read and Newns \cite{read83}, we take advantage of the local U(1) gauge
transformation of the lagrangian ${\cal L}^{\prime}(\tau)$

\[
\Phi _{i}\rightarrow r_{i}\exp (i\theta _{i}) 
\]
\[
f_{i}\rightarrow f_{i}^{\prime }\exp (i\theta _{i}) 
\]
\[
\lambda _{i}\rightarrow \lambda _{i}^{^{\prime }}+i~\partial \theta
_{i}/\partial \tau 
\]
We use the radial gauge in which the modulus of both fields $\Phi _{i}$ and $%
\Phi _{i}^{\ast }$ are the radial field $r_{i}$, and their phase $\theta
_{i} $ (via its time derivative) is incorporated into the Lagrange
multiplier $\lambda _{i}$ which turns out to be a field. Use of the radial
instead of the cartesian gauge bypasses the familiar complications of
infrared divergences associated with unphysical Goldstone bosons. We let the
fields fluctuate away from their saddle-point values : $r_{i}=r_{0}+\delta
r_{i}$, $\lambda _{i}=\lambda _{0}+\delta \lambda _{i}$, $\bf{%
\xi }_{f_{i}}=\delta \mbox{\boldmath{$\xi$}}_{f_{i}}$ and $\bf{\xi 
}_{c_{i}}=\delta \mbox{\boldmath{$\xi$}}_{c_{i}}$. After integrating out the
Grassmann variables in the partition function in Equation (\ref{eq6}), we get 

\begin{equation}
\label{eq12}
Z=\int {\cal D}r_{i}{\cal D} \lambda _{i}{\cal D}\mbox{\boldmath{$\xi$}}_{f_{i}}{\cal D}\mbox{\boldmath{$\xi$}}_{c_{i}}\exp
[-S_{eff}]
\end{equation}
where the effective action is

\[
S_{eff}=-\sum_{k,i\omega _{n}}Ln~Det{\bf G}^{-1}({\bf{k}},i\omega
_{n})+\beta ~[~J_{C}\sum_{i}r_{i}^{2}+J_{S}\sum_{i}\mbox{\boldmath{$\xi$}}%
_{f_{i}} \cdot \mbox{\boldmath{$\xi$}}_{c_{i}}+N_{S}(\mu n_{c}-\lambda _{0})]
\]
with : 
\[
\left[ {\bf G}^{-1}(i\omega _{n})\right] _{ij}^{\sigma \sigma ^{\prime }}=\left( 
\begin{array}{cc}
\lbrack (-i\omega _{n}-\mu )\delta _{ij}-t_{ij}]\delta _{\sigma \sigma
^{\prime }}-J_{S}i\mbox{\boldmath{$\xi$}}_{f_{i}}.\mbox{\boldmath{$\tau$}}%
_{\sigma \sigma ^{\prime }}\delta _{ij} & (\sigma _{0}+J_{C}\delta

r_{i})\delta _{\sigma \sigma ^{\prime }}\delta _{ij} \\ 
(\sigma _{0}+J_{C}\delta r_{i})\delta _{\sigma \sigma ^{\prime }}\delta _{ij}
& [-i\omega _{n}+\varepsilon _{f}+\delta \lambda _{i}]\delta _{\sigma \sigma
^{\prime }}\delta _{ij}-J_{S}i\mbox{\boldmath{$\xi$}}_{c_{i}}.\mbox{\boldmath
{$\tau$}}_{\sigma \sigma ^{\prime }}\delta _{ij}
\end{array}
\right) 
\]
Expanding up to the second order in the Bose fields, one obtains the
gaussian corrections $S_{eff}^{(2)}$ to the saddle-point effective action 
\bea
\label{eq13}
S_{eff}^{(2)} &=&\frac{1}{\beta} \sum_{{\bf{q}},i\omega_{\nu}}[\left( 
\begin{array}{cc}
\delta r, & \delta \lambda 
\end{array}
\right) {\bf D}_{C}^{-1}({\bf{q}},i\omega_{\nu})\left( 
\begin{array}{c}
\delta r \\ 
\delta \lambda 
\end{array}
\right)  +\left( 
\begin{array}{cc}
\delta \xi _{f}^{z}, & \delta \xi _{c}^{z}
\end{array}
\right) {\bf D}_{S}^{\Vert -1}({\bf{q}},i\omega_{\nu})\left( 
\begin{array}{c}
\delta \xi _{f}^{z} \\ 
\delta \xi _{c}^{z}
\end{array}
\right)  \nonumber \\
&+&\left( 
\begin{array}{cc}
\delta \xi _{f}^{+}, & \delta \xi _{c}^{+}
\end{array}
\right) {\bf D}_{S}^{\bot -1}({\bf{q}},i\omega_{\nu})\left( 
\begin{array}{c}
\delta \xi _{f}^{-} \\ 
\delta \xi _{c}^{-}
\end{array}
\right) +\left( 
\begin{array}{cc}
\delta \xi _{f}^{-}, & \delta \xi _{c}^{-}
\end{array}
\right) {\bf D}_{S}^{\bot -1}({\bf{q}},i\omega_{\nu})\left( 
\begin{array}{c}
\delta \xi _{f}^{+} \\ 
\delta \xi _{c}^{+}
\end{array}
\right)] 
\eea
where the boson propagators split into the following charge and longitudinal spin parts 
\be
\label{eq14}
{\bf D}_{C}^{-1}({\bf{q}},i\omega_{\nu})=\left( 
\begin{array}{cc}
J_{C}[1-J_{C}(\overline{\varphi }_{2}({\bf{q}},i\omega_{\nu})+\overline{\varphi }_{m}({\bf{q}},i\omega_{\nu}))] & 
-J_{C}\overline{\varphi }_{1}({\bf{q}},i\omega_{\nu}) \\ 
-J_{C}\overline{\varphi }_{1}({\bf{q}},i\omega_{\nu}) & -\overline{\varphi }_{ff}({\bf{q}},i\omega_{\nu})
\end{array}
\right) 
\ee
\[
{\bf D}_{S}^{\Vert -1}({\bf{q}},i\omega_{\nu})=\left( 
\begin{array}{cc}
J_{S}^{2}\varphi _{ff}^{\Vert }({\bf{q}},i\omega_{\nu}) & J_{S}[1+J_{S}\varphi _{cf}^{\Vert }({\bf{q}},i\omega_{\nu})]
\\ 
J_{S}[1+J_{S}\varphi _{fc}^{\Vert }({\bf{q}},i\omega_{\nu})] & J_{S}^{2}\varphi _{cc}^{\Vert }({\bf{q}},i\omega_{\nu})
\end{array}
\right) 
\]
and equivalent expression for the transverse spin part ${\bf D}_{S}^{\bot -1}({\bf{q}},i\omega_{\nu})$. The
expression of the different bubbles are given in the appendix. The charge boson propagator
${\bf D}_{C} ({\bf{q}},i\omega_{\nu})$ associated to the Kondo effect is equivalent to that obtained in
the $1/N$ expansion theories. The originality of the approach is to simultaneously derive
the spin propagator ${\bf D}_{S}^{\Vert -1}({\bf{q}},i\omega_{\nu})$ and ${\bf D}_{S}^{\bot -1}({\bf{q}},i\omega_{\nu})$ associated to the spin fluctuation effects.
Note that in the magnetically-disordered phase, the charge and spin contributions in $S_{eff}$
are totally decoupled. 

\subsection{\protect\bigskip Dynamical spin susceptibility}

Next step is to consider the dynamical spin susceptibility. For that
purpose, we study the linear response $M_{f}$ to the source-term $-2%
{\bf{S}}_{f}.\bf{B}$ (we consider $\bf{B}$
colinear to the $\bf{z}$-axis). The effect on the partition
function expressed in Equation (\ref{eq6}) is to change the hamiltonian $H^{\prime }_{J}(\tau )$ to $%
H^{\prime B}_{J}(\tau )$  
\be
\label{eq15}
H^{\prime B}_{J}(\tau )=\sum_{i\sigma \sigma ^{\prime
}}\left( 
\begin{array}{cc}
c_{i\sigma }^{\dagger}, & f_{i\sigma }^{\dagger}
\end{array}
\right) \left( 
\begin{array}{cc}
-J_{S}i\mbox{\boldmath{$\xi$}}_{f_{i}}\cdot {\mbox{\boldmath{$\tau$}}}_{\sigma \sigma ^{\prime
}} & J_{C}\Phi
_{i}^{\ast }\tau^{0}_{\sigma \sigma ^{\prime
}} \\ 
J_{C}\Phi _{i}\tau^{0}_{\sigma \sigma ^{\prime
}} & \sum\limits_{\alpha =x,y,z}(-J_{S}i\xi
_{c_{i}}^{\alpha }-B\delta _{\alpha z}).\tau^{\alpha}_{\sigma \sigma ^{\prime
}}
\end{array}
\right) \left( 
\begin{array}{c}
c_{i\sigma ^{\prime }} \\ 
f_{i\sigma ^{\prime }}
\end{array}
\right) +  J_{C}\sum_{i}\Phi _{i}^{\ast }\Phi _{i}+J_{S}\sum_{i}%
\mbox{\boldmath{$\xi$}}_{f_{i}}.\mbox{\boldmath{$\xi$}}_{c_{i}}
\ee
Introducing the change of variables $\xi _{c_{i}}^{\alpha }=\xi
_{c_{i}}^{\alpha }-iB/J_{S}$, we connect the f magnetization and the ff
dynamical spin susceptibility to the Hubbard Stratonovich fields $%
\mbox{\boldmath{$\xi$}}_{f_{i}}$ 
\[
M_{f}^{z}=-\frac{1}{\beta }\frac{\partial LnZ}{\partial B_{z}}=i\left\langle
\xi _{f_{i}}^{z}\right\rangle 
\]
\begin{equation}
\label{eq16}
\chi _{ff}^{\alpha \beta }=-\frac{1}{\beta }\frac{\partial ^{2}LnZ}{\partial
B^{\alpha }\partial B^{\beta }}=-
\left\langle \xi _{f_{i}}^{\alpha} \xi _{f_{i}}^{\beta}\right\rangle + 
\left\langle  \xi _{f_{i}}^{\alpha} \right\rangle  \left\langle \xi _{f_{i}}^{\beta}\right\rangle 
\end{equation}
Using the expression (\ref{eq14}) fot the boson propagator ${\bf D}_{S}^{\Vert -1}({\bf q})$, we get for the longitudinal spin susceptibility
\begin{equation}
\label{eq17}
\chi _{ff}^{\Vert }({\bf{q}},i\omega_{\nu})=\frac{\varphi _{ff}^{\Vert }({\bf{q}},i\omega_{\nu})}{1-J_{S}^{2}[\varphi
_{ff}^{\Vert }({\bf{q}},i\omega_{\nu})\varphi _{cc}^{\Vert }({\bf{q}},i\omega_{\nu})-\varphi _{fc}^{\Vert 2}({\bf{q}},i\omega_{\nu})-\frac{2}{%
J_{S}}\varphi _{fc}^{\Vert }({\bf{q}},i\omega_{\nu})]}
\end{equation}
and equivalent expression for the transverse spin susceptibility $\chi
_{ff}^{\bot }({\bf{q}},i\omega_{\nu})$. The diagrammatic representation of Equation (\ref{eq17}) is reported in Figure 2. The different bubbles $\varphi _{ff} ({\bf{q}},i\omega_{\nu})$, $\varphi _{cc} ({\bf{q}},i\omega_{\nu})$
and $\varphi _{fc} ({\bf{q}},i\omega_{\nu})$ are evaluated from the expressions of the Green's
functions 
\be
\label{eq18}
G_{ff}({\bf{k}},i\omega_{n})=u_{k}^{2}G_{\alpha \alpha }({\bf{k}},i\omega_{n})+v_{k}^{2}G_{\beta
\beta }({\bf{k}},i\omega_{n})
\ee
\[
G_{cc}({\bf{k}},i\omega_{n})=v_{k}^{2}G_{\alpha \alpha }({\bf{k}},i\omega_{n})+u_{k}^{2}G_{\beta
\beta }({\bf{k}},i\omega_{n})
\]
\[
G_{cf}({\bf{k}},i\omega_{n})=G_{fc}({\bf{k}},i\omega_{n})=-u_{k}v_{k}[G_{\alpha \alpha}
({\bf{k}},i\omega_{n})-G_{\beta \beta }({\bf{k}},i\omega_{n})]
\]
where $G_{\alpha \alpha }({\bf k},i\omega_n)$ and $G_{\beta \beta }({\bf k},i\omega_n)$
are the Green's functions associated to the eigenstates $\alpha _{k\sigma }^{\dagger}|0>$%
and $\beta _{k\sigma }^{\dagger}|0>$. In the low frequency limit, one can easily check that the
dynamical spin susceptibility may be written as

\begin{equation}
\label{eq19}
\chi _{ff}({\bf{q}},i\omega_{\nu})=\frac{\chi _{\alpha \alpha }({\bf{q}},i\omega_{\nu})+\overline{\chi}
_{\alpha \beta }({\bf{q}},i\omega_{\nu})}{1-J_{S}^{2}\chi _{\alpha \alpha }({\bf{q}},i\omega_{\nu})%
\overline{\chi} _{\alpha \beta }({\bf{q}},i\omega_{\nu})}
\end{equation}
for both the longitudinal and the transverse parts. 
\[
\chi _{\alpha \alpha }({\bf{q}},i\omega_{\nu})=\frac{1}{\beta }\sum\limits_{k}\frac{%
n_{F}(E_{k}^{-})-n_{F}(E_{k+q}^{-})}{i\omega _{\nu }-E_{k+q}^{-}+E_{k}^{-}}
\]
\[
\overline{\chi} _{\alpha \beta }({\bf{q}},i\omega_{\nu})=\frac{1}{\beta }\sum%
\limits_{k}(u_{k}^{2}v_{k+q}^{2}+v_{k}^{2}u_{k+q}^{2})\frac{%
n_{F}(E_{k}^{-})-n_{F}(E_{k+q}^{+})}{i\omega _{\nu }-E_{k+q}^{+}+E_{k}^{-}}
\]

\begin{figure}
\centerline{\psfig{file=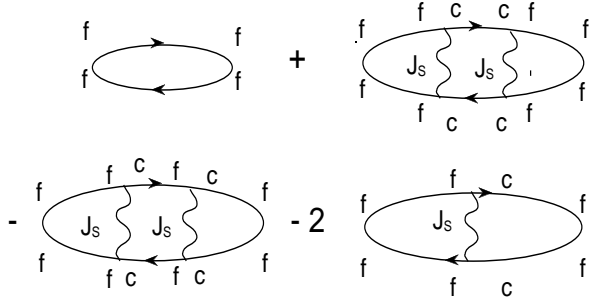,height=5cm,width=8cm}}
\caption{Diagrammatic representation of Equation (\ref{eq27}) for the dynamical spin susceptibility $\chi _{ff}(\bf{q},\omega)$.}
\label{fig2}
\end{figure}

\subsection{\protect\bigskip Physical discussion}
 
From Equation (\ref{eq19}), one can see that the dynamical spin susceptibility is made
of two contributions $\chi _{intra}({\bf{q}},i\omega_{\nu})$ and $\chi _{inter}({\bf{q}},i\omega_{\nu})$

\be
\label{eq20}
\chi _{ff}({\bf{q}},i\omega_{\nu})=\chi _{intra}({\bf{q}},i\omega_{\nu})+\chi _{inter}({\bf{q}},i\omega_{\nu}) 
\ee

with

\be
\label{eq21}
\chi _{intra}({\bf{q}},i\omega_{\nu})=\frac{\chi _{_{\alpha \alpha }}({\bf{q}},i\omega_{\nu})}{1-J_{S}^{2}\chi
_{_{\alpha \alpha }}({\bf{q}},i\omega_{\nu})\overline{\chi }_{\alpha \beta }({\bf{q}},i\omega_{\nu})}
\ee

\be
\label{eq22}
\chi _{inter}({\bf{q}},i\omega_{\nu})=\frac{\overline{\chi}_{\alpha \beta }({\bf{q}},i\omega_{\nu})}{%
1-J_{S}^{2}\chi _{_{\alpha \alpha }}({\bf{q}},i\omega_{\nu})\overline{\chi }_{\alpha \beta }({\bf{q}},i\omega_{\nu})}
\ee

$\chi _{intra}({\bf{q}},i\omega_{\nu})$and $\chi _{inter}({\bf{q}},i\omega_{\nu})$ respectively represent the
renormalized particle-hole pair excitations within the lower $\alpha $ band, and
from the lower $\alpha $ to the upper $\beta $ band. The latter expression is
reminiscent of the behaviour proposed by Bernhoeft and Lonzarich \cite{bernhoeft} to explain
the neutron scattering observed in $UPt_{3}$ with the existence of both a "slow" 
and a "fast" component in $\chi^{"}(\bf{q},\omega)/{\omega}$ due to spin-orbit 
coupling. Also in a phenomenological way, the same type of feature has been suggested 
in the duality model developed by Kuramoto and Miyake \cite{kuramoto90}. To our knowledge, the proposed approach
provides the first microscopic derivation from the Kondo lattice model of
such a behaviour.
The bare intraband susceptibility $\chi _{_{\alpha \alpha }}({\bf{q}},\omega )$ is well approximated by a lorentzian

\be
\label{eq23}
\chi _{_{\alpha \alpha }}^{-1}({\bf{q}},\omega )=\rho _{\alpha \alpha }({\bf{q}})^{-1} \left( 1-i\frac{%
\omega }{\Gamma _{0}({\bf{q}})}\right) 
\ee

where $\rho _{\alpha \alpha }=\chi _{_{\alpha \alpha }}^{\prime}({\bf{q}},0)$ and $\Gamma _{0}(\bf{q})$ is  the relaxation rate of order $\left| y_{F}\right|=T_{K}$. This
corresponds to the Lindhard continuum of the intraband particle-hole pair 
excitations $\chi _{_{\alpha \alpha }}^{"}(q,\omega)\neq 0$ as reported in Figure 3. 
In the same way, one can schematize the low-frequency behavior $(\omega<<\omega _{0}({\bf{q}})$ of the bare interband susceptibility by

\be
\label{eq26}
\overline{\chi }_{\alpha \beta }^{\prime,-1}({\bf{q}},\omega )=\rho _{\alpha \beta}({\bf{q}})^{-1}\left( 1-\frac{%
\omega }{\omega _{0}({\bf{q}})}\right) 
\ee

where $\rho _{\alpha \beta}=\overline{\chi }_{_{\alpha \beta}}^{\prime}({\bf{q}},0)$ and $\omega _{0}(\bf{q})$ is a characteristic frequency-scale of the interband transitions. The value of $\omega _{0}(\bf{q})$
is strongly structure-dependent. In the simple case of a cubic band 
structure $\epsilon_{k}=-2t(\cos k_{x}+\cos k_{y}+\cos k_{z})$ 
(tight-binding scheme including nearest-neighbor hopping), we find 
a weakly wavevector dependent frequency around 
${\bf{q}}={\bf{Q}}$ of order of 
$\omega _{0}=2\left| y_{F}\right| /\left( \rho _{0}J_{C}\right)$. 
The latter result does not stand for more complicated band structures
as obtained by de Haas-van Alphen studies combined with band structure 
calculations in heavy-Fermion compounds. 
In the following, we will leave $\omega _{0}(\bf{q})$ as a parameter.
Figure 3 reports the continuum of interband particle-hole excitations 
$\overline{\chi}_{\alpha \beta }"\neq 0$. Due to the presence 
of the hybridization gap in
the density of states, the latter continuum displays a gap equal to $2\sigma
_{0}$, the value of the direct gap at ${\bf{q}}={\bf{0}}$%
, and $2\left| y_{F}\right| $, the value of the indirect gap at 
${\bf{q}}={\bf{Q}}$ (close to $k_{F}$). More precisely,
we have

\be
\label{eq27}
\overline{\chi }_{\alpha \beta }"({\bf{0}},\omega )=4\rho _{0}%
\frac{\sigma _{0}^{2}}{\omega \sqrt{\omega ^{2}-4\sigma _{0}^{2}}}\text{ at }%
2\sigma _{0}<\omega <D
\ee

\[
\overline{\chi }_{\alpha \beta }"({\bf{Q}},\omega )=2\rho _{0}%
\frac{1}{1+\omega ^{2}/(2\sigma _{0})^{2}}\text{ at }2\left| y_{F}\right|
<\omega <2D
\]

\begin{figure}
\centerline{\psfig{file=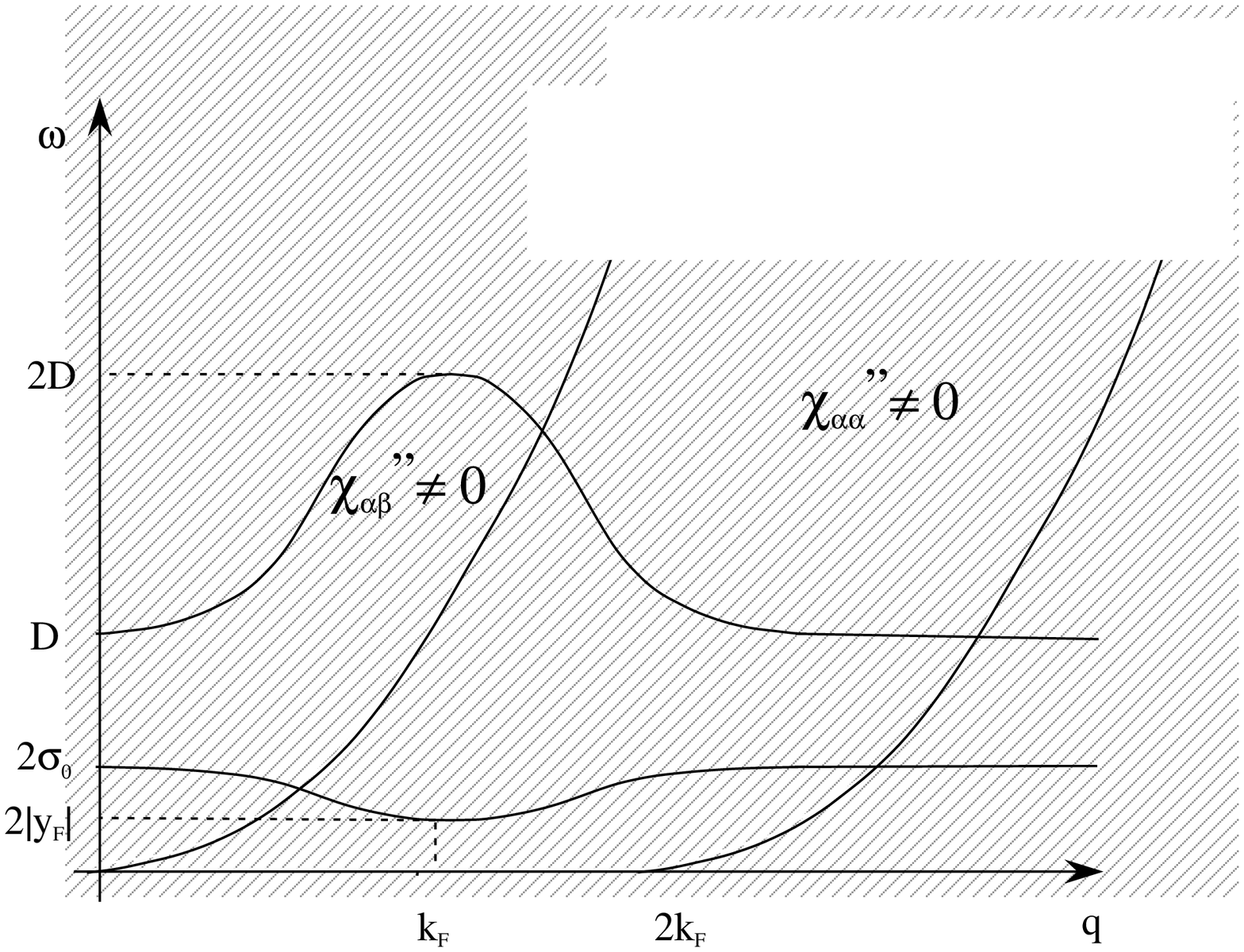,height=5cm,width=8cm}}
\caption{Continuum of the intra- and interband electron-hole pair excitations 
$\chi _{_{\alpha \alpha}}^{"}(q,\omega)\neq 0$ and $\chi _{_{\alpha \beta}}^{"}(q,\omega)\neq 0$.
Note the presence of a gap in the interband transitions equal to the indirect gap of value 
$2\left| y_{F}\right|$ at $q=k_F$, and to the direct gap of value $2\sigma _{0}$ at $q=0$.
}
\label{fig3}
\end{figure}

Far from the antiferromagnetic wavevector $\bf{Q}=(\pi,\pi,\pi)$, $\chi_{ff} ({\bf{q}},\omega )$
is dominated by the intraband transitions. In the low frequency limit, the frequency dependence of $\chi _{intra}^{"}(\bf{q},\omega)$ can be approximate to a lorentzian 

\be
\label{eq24}
\chi_{ff}^{"} ({\bf{q}},\omega ) \approx \chi _{intra}^{"}({\bf{q}},\omega)=\omega  \frac{\chi
_{intra}^{^{\prime }}({\bf{q}})\Gamma _{intra}({\bf{q}})}{\omega ^{2}+{\Gamma
_{intra}({\bf{q}})}^{2}} 
\ee

with

\be
\label{eq25}
\Gamma _{intra}({\bf{q}})=\Gamma _{0}({\bf{q}})(1-I({\bf{q}})) 
\ee

\[
\chi _{intra}^{^{\prime }}({\bf{q}})=\frac{\rho _{\alpha \alpha }({\bf{q}})}{(1-I({\bf{q}}))} 
\]

$I({\bf{q}})=J_{S}^{2}\chi_{\alpha \alpha}^{\prime}({\bf{q}},0) \overline{\chi }_{\alpha \beta }^{\prime}
({\bf{q}},0)$. One has: $\chi_{\alpha \alpha}^{\prime}({\bf{0}},0)=\rho_{\alpha \alpha }({\bf{0}})=\rho(E_F)$ and $\chi_{\alpha \beta}^{\prime}({\bf{0}},0)=\rho_{0}$.
The contribution expressed in equation (\ref{eq24}) is
consistent with the standard Fermi liquid theory. Note that the product $\Gamma _{intra}({\bf{q}})\chi _{intra}^{^{\prime }}({\bf{q}})=\rho_{\alpha\alpha}({\bf{q}})\Gamma_{0}({\bf{q}})$ is independent
of I. 

\bigskip

Oppositely, at the antiferromagnetic wavevector $\bf {Q}$, $\chi_{ff} ({\bf{q}},\omega )$
is driven by the interband contribution and we get

\be
\label{eq28}
\chi_{ff}^{"} ({\bf{Q}},\omega ) \approx \chi _{inter}^{"}({\bf{Q}},\omega )=\omega \frac{I\chi
_{inter}^{\prime }\Gamma _{inter}}{(\omega -\omega _{\max })^{2}+\Gamma
_{inter}^{2}}
\ee

with

\be
\label{eq29}
\omega _{\max } =\omega _{0}(1-I)
\ee

\[
\Gamma _{inter} =\omega _{0}^{2}(1-I)/\Gamma _{0}
\]

\[
\chi _{inter}^{\prime } =\rho _{\alpha \beta }/(1-I)
\]

where  $\omega _{0}$, $\rho _{\alpha \beta }$, $\Gamma_{0}$ and $I$ are the
values of $\omega _{0}(\bf{q})$, $\rho _{\alpha \beta }(\bf{q})$
and $\Gamma_{0}(\bf{q})$ and $I(\bf{q})$ at $\bf{q}=\bf{Q}$.
The role of the interband transitions have already been pointed out in
previous works \cite{auerbach88}. However while the previous studies conclude to the presence
of an inelastic peak at finite value of the frequency related to the
hybridization gap whatever the interaction J is, we emphasize that the
renormalization of $\overline{\chi }_{\alpha \beta }({\bf{Q}},\omega )$ into $\chi _{inter}
({\bf{Q}},\omega )$ leads to a noteworthy renormalization of
the interband gap.
Due to the damping introduced by intraband transitions, $\chi _{inter}^{"}(\bf{Q},\omega )$
takes a finite value at frequency much smaller than the hybridization gap. Both the relaxation rate $%
\Gamma _{inter}$ vanishes and the susceptibility $\chi _{inter}^{\prime }$
diverges at the antiferromagnetic transition with again the product $\Gamma_{inter} \chi_{inter}^{\prime }$
independent of $I$. Remarkably, the value $%
\omega _{\max }$ of the maximum of $\chi _{inter}^{"}(\bf{Q},\omega )/{\omega}$
is at the same time pushed to zero. This excitation can be analyzed as an excitonic mode which softens at the magnetic transition. 
Such a behaviour has
been effectively observed in $Ce_{1-x}La_{x}Ru_{2}Si_{2}$ \cite{raymond} with a reduction of $\Gamma _{inter}$
and $\omega _{\max }$ respectively by a factor 4 and 6 when x goes from 0 to 0.075 so when
getting closer to the magnetic instability occuring at $x=0.08$. It is likely that this mode is called to play a role in the critical phenomena observed near the magnetic transition.
\bigskip 

\section{\protect\bigskip The Supersymmetric approach}
Traditionally, the spin is described either in fermionic or bosonic representation. If the former representation, used for instance in the $1/N$ expansion of the Anderson or the Kondo lattice models, appears to be well adapted for the description of the Kondo effect, it is also clear that the bosonic representation lends itself better to the study of local magnetism. Quite obviously the physics of  heavy-Fermions is dominated by the duality between Kondo effect and localized moments. This constitutes the motivation to introduce a new approach to the Kondo lattice model (KLM) which relies on an original representation of the impurity spin $1/2$ in which the different degrees of freedom are represented by fermionic as well as bosonic variables. The former are believed to describe the Fermi liquid excitations while the latter account for the residual spin degrees of freedom.
\par
In order to include the Fermi liquid excitations as well as the residual spin degrees of freedom, the proposition is to enlarge the representation of the spin operator as follows
\be
S^{a} =  
\sum_{\sigma\sigma'}{b_{\sigma}^{\dagger}\tau^{a}_{\sigma\sigma'}b_{\sigma'}+f^\dagger_{\sigma}\tau^{a}_{\sigma\sigma'}f_{\sigma'}}
 = S^{a}_{b}+S^{a}_{f}
\ee
where $b^\dagger_\sigma$ and $f^\dagger_\sigma$ are respectively bosonic and fermionic creation operators and $\tau^a$ $(a = \left(+,-,z\right))$ are Pauli matrices. Eq.(1) corresponds to a mixed fermionic-bosonic representation between Schwinger bosons and Abrikosov pseudo-fermions. To
restrict the dimension of the Hilbert space to two, we introduce the following local constraints
\be 
\displaystyle{n_f+n_b = 1}  
\ee

The constraint restricts the Hilbert space to the two states of the
form: $|\upar\rangle = (X b^{\dagger}_{\upar}+Y f^{\dagger}_{\upar})|0\rangle$,
 $|\dn\rangle = (X b^{\dagger}_{\dn}+Y f_{\dn}^{\dagger})|0\rangle$ where $X^2+Y^2=1$ to guarantee the state normalization to $1$ and $\left| 0 \right\rangle$ represents the vacuum of particles:  $b_\sigma \left| 0 \right\rangle =f_\sigma \left| 0 \right\rangle =0$. $X$ and $Y$ are parameters controlling the weight of boson and fermion statistics in the representation: they will be fixed later on by the dynamics. The constraint can be viewed as a charge conservation of the following $SU(1|1)$ fermion-boson rotation symmetry leaving the spin operator invariant
\be
\left( f'^\dagger_\sigma, b'^\dagger_\sigma \right) = \left( f^\dagger_\sigma, b^\dagger_\sigma \right) V^\dagger
\ee
where $V^\dagger$ is an unitary supersymmetric matrix ($V V^\dagger=V^\dagger V=1$). One can easily check that
the representation satisfies the standard rules of $SU(2)$ algebra: $|\upar\rangle$ and  $|\dn\rangle$ are eigenvectors of $S^2$ and $S^z$ with eigenvalues $3/4$ and $\pm 1/2$ respectively, $\left[S^+,S^- \right]=2 S^z$ and $\left[S^z,S^{\pm}\right]=\pm S^{\pm}$ provided that the local constraints expressed in Eq.(2) are satisfied.

\par

In the representation introduced before, the partition function of the three-dimensional KLM can be written as the following path integral 
\be \begin{array}{l}
Z=\sum_{n=1,2}  \int {\cal D}c_{i\sigma}{\cal D}f_{i\sigma}{\cal D}b_{i\sigma}
d \nu_{i} d \lambda_{i}^{(n)} 
\exp  ( - \int_{0}^{\beta} d\tau \\  \nonumber
({\cal L(\tau)} + {\cal H}  
+\sum_{i} \nu_{i} P_{i} + \sum_{i} \lambda_{i}^{(n)} Q_{i}^{(n)}) )  \  \\  \\
\begin{array}{ll} 
\mbox{with} & {\cal L(\tau)}=\sum_{i\sigma} (c_{i\sigma}^\dagger\partial_{\tau}c_{i\sigma}+f_{i\sigma}^{\dagger} \partial_{\tau}f_{i\sigma} +b_{i\sigma}^{\dagger}\partial_{\tau}b_{i\sigma}) \\ \\
\mbox{and} &  
{\cal H}=\sum_{k\sigma} {\ek c_{k\sigma}^{\dagger}c_{k\sigma}}  \\
& + J\sum_{i} ({\bf S}_{f_{i}}+{\bf S}_{b_{i}}).{\bf s}_{i}  -\mu\sum_{i}{n_{c_{i}}} \  
         \end{array} \end{array} \ee 
Note the presence of two terms in $Z$ coming from the contributions of the states satisfying $Q_{i}^{(n)}=S^{z}_{i}+(-1)^{n}/2=0$ respectively for $n=1$ and $2$.  The time-independent Lagrange multipliers $\lambda_{i}^{(n)}$ and $\nu_{i}$ are introduced to enforce the local constraints  $Q_{i}^{(n)}=0$ and $P_{i}=n_{f_{i}}+n_{b_{i}}-1=0$.
Performing a Hubbard-Stratonovich transformation and neglecting the space and time dependence of the fields in a self-consistent saddle-point approximation, we have 
\be
\label{sadpoint} \begin{array}{l}
Z=\sum_{n=1,2}  \int{d\eta d\eta^{*}{\cal C}_n\left(\sigma_0, \lambda_0, \eta ,\eta*\right)Z_{n}(\eta,\eta^{*})}  \\ \\
Z_n (\eta,\eta^{*})=\sum_{\sigma} \int{\cal D}c_{i\sigma}{\cal D}f_{i\sigma}{\cal D}b_{i\sigma}  \exp \left( - \int_0^\beta d \tau ( {\cal L(\tau)} + {\cal H'}_{n\sigma} ) \right)  \  \\ \\
\begin{array}{ll}
\mbox{with} 
& {\cal H'}_{n\sigma} = \sum_{k}{(
f^{\dagger}_{k\sigma}, c^{\dagger}_{k\sigma}, b^{\dagger}_{k\sigma})
H_{0}^{n\sigma}\left(
\begin{array}{c}
f_{k\sigma} \\
c_{k\sigma} \\
b_{k\sigma}
\end{array}
\right)}  \  \\    
& H_{0}^{n\sigma}= \left(
\begin{array}{ccc}
\ef+\sigma \lambda_{0}^{(n)}/2 & \sigma_{0} & 0 \\
\sigma_{0} & \ek & \eta \\
0 & \eta^{*} & \ef+\sigma \lambda_{0}^{(n)}/2
\end{array}
\right) \,
\end{array} 
\end{array} 
\ee
where ${\cal C}_n\left(\sigma_0, \lambda_0, \eta ,\eta*\right)$ is an integration constant. $\ef$ and $\lambda_{0}^{(n)}$ are the saddle-point values of the Lagrange multipliers $\nu_{i}$ and $\lambda_{i}^{(n)}$.
Note the presence of a Grassmannian coupling $\eta$ between $c_{i \sigma}$ and $b_{i \sigma}$, in addition to the usual coupling $\sigma_0$ between $c_{i \sigma}$ and $f_{i \sigma}$ responsible for the Kondo effect. In the following, 
$H_0$ is indifferently used for any $H_{0}^{n\sigma}$. $H_0$ is of the
type $\left( \begin{array}{cc} a & \sigma \\ \rho & b \end{array} \right)$ in which $ a$,$b$ ($\rho$,$\sigma$) are matrices consisting of commuting (anticommuting) variables. Note the supersymmetric structure of the matrix 
$H_0$ similar to the supermatrices appearing in the theory of disordered
metals~\cite{efetov}. 

$H_0$ being hermitian, the matrix $U^\dagger$ transforming the original basis $\psi^\dagger = \left( f^\dagger, c^\dagger, b^\dagger \right)$ to the basis of eigenvectors $\Phi^\dagger = \left( \alpha^\dagger, \beta^\dagger, \gamma^\dagger \right)$ is unitary ($U U^\dagger=U^\dagger U=1$). $\Phi^\dagger 
= \psi^\dagger U^\dagger$ with $U^\dagger$ a supersymmetric matrix. 
$\alpha^\dagger$ and $\beta^\dagger$ are the fermionic eigenvectors whose eigenvalues, determined from $
\det\left[(a-E)-\sigma(b-E)^{-1}\rho\right]=0 $, are
$$
E_{\mp} =
\frac{(\ek+\ef)\mp \sqrt{(\ek-\ef)^{2}+4(\sigma_{0}^{2}+\eta\eta^{*})}
}{2} \ .
$$
$\gamma^\dagger$ is the bosonic eigenvector whose eigenvalue, determined~ from $\det\left[(b-E)-\rho(a-E)^{-1}\sigma\right]=0$ is $E_\gamma=\ef$. 

In this scheme, $\sigma_0$ and $\lambda_0$ are slow variables that we determine by solving saddle-point equations, while $\eta$, $\eta^*$ are fast variables defined by a local approximation. As we will see,
the latter approximation incorporates part of the fluctuation effects. Indeed, performing the functional integration of Eq.(\ref{sadpoint}) over the fermion and boson fields~\cite{efetov} yields a superdeterminant ($SDet$) form written as follows
\par
\be
\begin{array}{l}
Z(\eta, \eta^*) = SDet(\partial_\tau + H_0) \ , \\ \\
   \begin{array}{ll}
\mbox{where} &  {\displaystyle SDet(\partial_\tau + H) = \frac{Det(G^{-1} - \sigma D \rho ) }{ Det( D^{-1})} }\ , \end{array} \\
\begin{array}{lll}
G^{-1}= \partial_\tau + a  & \mbox{and} & D^{-1} = \partial_\tau + b \ .
\end{array} \end{array} 
\ee

Expanding to second order in $\eta$, $\eta^*$ allows us to define the propagator $G_{\eta \eta^*}( {\bf k}, i \omega_n) $ associated to the Grassmann variable $\eta$ and hence the closure relation for $x_0^2 = \left \langle \eta \eta^* \right\rangle$
\be
\label{4}
{\displaystyle x_{0}^{2}=\frac{1}{\beta}\sum_{{\bf k},i\omega_{n}}G_{\eta \eta^*}( {\bf k}, i \omega_n)} \ ,
\ee 
$$ \begin{array}{ll}
\mbox{with} & {\displaystyle  G_{\eta \eta^*}( {\bf k}, i \omega_n) = \frac{J}{
\left[1- J\Pi_{cb}^{0}(\bf k,i\omega_{n})\right]} }\\
\mbox{and} & {\displaystyle \Pi_{cb}^0 = \frac{1}{\beta} \sum_{{\bf q}, i \omega_n} G_{cc} ({\bf k+q}, i \omega_n + i \omega_\nu) D( {\bf q}, i \omega_n) }\ .
\end{array}  $$

Contrary to \cite{gan} which assumes $x_0^2=0$ leading to a two-fluid model description, the closure equation Eq.(\ref{4}) defines a finite $x_0^2$. This parameter $x_0^2$ plays a major role in controlling the relative weights of fermion and boson statistics. It is directly connected to the X and Y
parameters introduced in the initial representation of the states : $X^2=x_0^2/(\sigma_0^2+x_0^2)$ and $Y^2=\sigma_0^2/(\sigma_0^2+x_0^2)$.

The resolution of the saddle-point equations, keeping the number of particles conserved, leads to
\be 
\label{5} 
\begin{array}{c}
{\displaystyle y_{F}= -D \exp \left[ -1/(2J\rho_{0}) \right] } \ , \\
{\displaystyle 1= \frac{2 \rho_0 ( \sigma_0^2 + x_0^2)}{- y_F} }\ , \\
{\displaystyle \mu = - \frac{(\sigma_0^2 + x_0^2)}{D} }\ , \\
{\displaystyle \lambda_{0}^{(1)}=\lambda_{0}^{(2)}=0 }\ ,   
\end{array} 
\ee 
where $y_{F}=\mu-\ef$ and $\rho_0 = 1/2D$ is the bare density of states of conduction electrons. From that set of equations, we find : $\ef =0$.

The resulting spectrum of energies is schematized in Figure~\ref{fig1}. At zero temperature, only the lowest band $\alpha$ is filled with an enhancement of the density of states at the Fermi level (and hence of the mass) unchanged from the standard slave-boson theories ${\displaystyle\frac{\rho(E_F)}{\rho_0}=1+\frac{(\sigma_0^2 + x_0^2)}{y_F^2} = 1+ \frac{D}{(-y_F)} \gg 1 }$. This large mass enhancement is related to the flat part of the $\alpha$ band associated with the formation of the Abrikosov-Suhl resonance pinned at the Fermi level. While this feature was already present in the purely fermionic description, it is to be noted that the formation of a dispersionless bosonic band within the hybridization gap is an entirely new result of the theory.

\begin{figure}
\centerline{\psfig{file=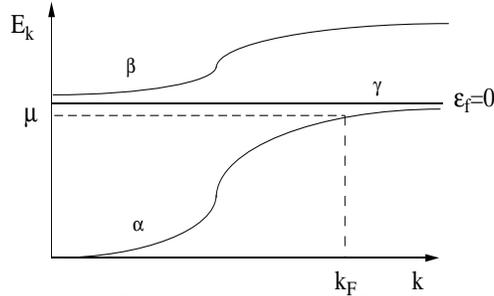,height=4cm,width=7cm}}
\caption{Sketch of energy versus wave number $k$ for the three bands $\alpha$, $\beta$, $\gamma$ resulting of the diagonalization of supersymmetric $H_0$.}
\label{fig1}
\end{figure}

The relative weight of boson and fermion statistics in the spin representation is related to $x_0^2$ : $n_b/n_f=x_0^2/{\sigma_0^2}$. It is then interesting to
follow the J-dependence of $x_0^2$ as determined by the closure equation (\ref{4}). The result is reported in Fig.~\ref{fig2}. This bell-shaped curve can be interpreted in the light of the exhaustion principle mentioned in the introduction. In the
limit of large J, the Kondo temperature-scale $T_K=D exp[-1/(2J\rho_0)]$ is of order of the bandwidth. One then expects a complete Kondo screening as can be checked by remarking that the weight of c in the $\alpha$ quasiparticle at the Fermi level (noted $v_{k_F}^2$) just equals the added weights of f and b at the Fermi level (respectively noted $u_{k_F}^2$ and $\rho_1^2$): $v_{k_F}^2/(u_{k_F}^2+\rho_1^2)=y_F^2/(\sigma_0^2+x_0^2)=1$. The Kondo effect being complete in that limit, there is no residual unscreened spin degrees of freedom: it is then natural to derive a zero value of $x_0^2$ (and hence of $n_b$). The opposite limit at small J corresponds to the free case of uncoupled impurity spins and conduction electrons. It also leads to: $x_0^2=0$. The finite value of $x_0^2$ between these two limits with a maximum reflects the incomplete Kondo screening effect in the Kondo lattice, the unscreened spin degrees of freedom being described by bosons. 

\begin{figure}
\centerline{\psfig{file=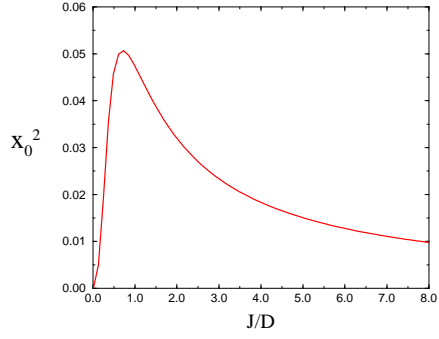,height=5cm,width=6cm}}
\caption{$J/D$-dependence of the coupling $x_0^2= \protect \left\langle \eta \eta^* \protect \right \rangle$ fixing the relative weight of fermion and boson statistics. The unit on the vertical-axis is $D^2$.}
\label{fig2}
\end{figure}

Largely discussed in the litterature \cite{tsunetsugu}  is the question concerning the Fermi surface sum rule: do the localized spins of the Kondo lattice contribute to the counting of states within the Fermi surface or do they not? Depending on the answer, one expects large or small Fermi surfaces. The supersymmetric theory leads to a firm conclusion in favour of the former. One can check that the number of states within the Fermi surface is just equal to $n_c+n_b+n_f$, i.e. $n_c+1$. The Fermi surface volume includes a contribution of one atate per localized spin in addition to that of conduction electrons \cite{tsunetsugu,martin}. The latter conclusion appears sensible if one recalls that the KLM is an effective hamiltonian derived from the periodic Anderson model (PAM).

Let us now consider the response functions to some external fields namely the dynamical spin susceptibility $\chi^{ab}(\bf {q},\omega)$ and the frequency-dependent optical conductivity $\sigma ^{ab}(\omega )$ $(a,b=x,y,z)$. For that purpose, we introduce the Matsubara correlation functions associated with the operator ${\cal O}^{a}({\bf {q}},\tau)$: $\chi ^{ab}({\bf {q}},i\omega _{\nu })=\int_{0}^{\beta}d\tau \,\exp
^{i\omega _{\nu }\tau }\left\langle T_{\tau }
{\cal O}^{a}({\bf {q}},\tau)
{\cal O}^{b}(- {\bf {q}},0)\right\rangle$. 
The operator related to the spin-spin correlation function is the $a$-component of the spin expressed in the mixed representation introduced in the paper by
$S^{a}({\bf {q}})=\sum_{k,\sigma ,\sigma ^{\prime }}
{f_{k+q,\sigma }^{\dagger }\tau _{\sigma \sigma ^{\prime }}^{a}
f_{k,\sigma ^{\prime }}+b_{k+q,\sigma }^{\dagger }
\tau _{\sigma \sigma^{\prime }}^{a}b_{k,\sigma ^{\prime }}}$.
As usual, the dynamical spin susceptibility is then derived from the spin-spin correlation function by the analytical continuation $i\omega _{\nu }\rightarrow \omega +i0^{+}$.
In the same way, the operator related to the current-current correlation
function is the $a$-component of the c-current. In the case of a cubic lattice: $J^{a}_{c}({\bf {q}})=2\sum_{k,\sigma ,\sigma ^{\prime }}\sin k_{a} {c_{k+q,\sigma }^{\dagger }c_{k,\sigma }}$. The frequency-dependent optical conductivity is then obtained from the current-current correlation
function by the analytical continuation following:
$\sigma ^{ab}(\omega )=\left[ \chi ^{ab}({\bf {q}},\omega
 +i0^{+})-\chi ^{ab}({\bf {q}}, i0^{+})\right] \,/\,i\omega$. 

\begin{figure}
\centerline{\psfig{file=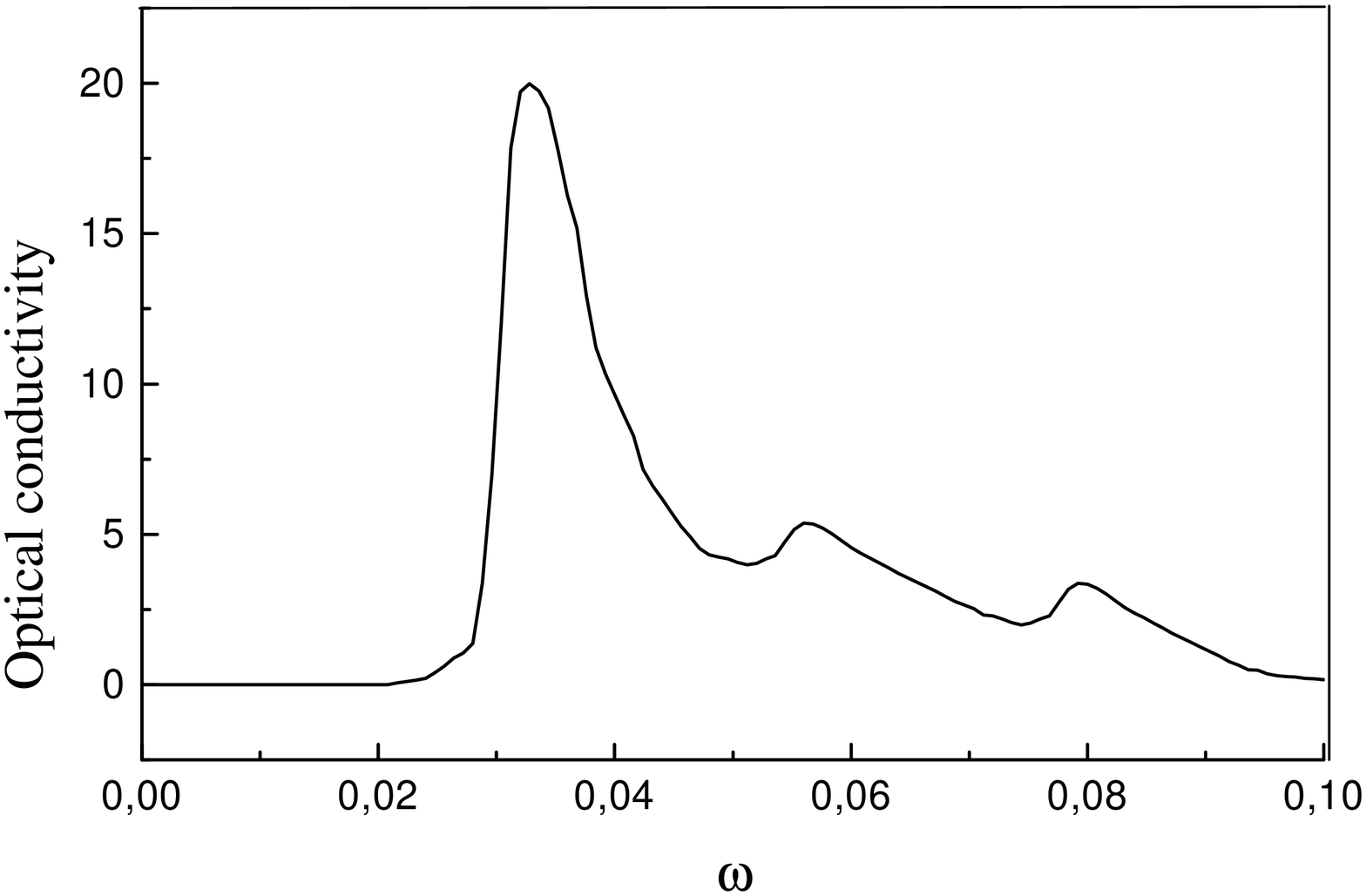,height=4cm,width=5cm}}
\caption{Frequency-dependence of the optical conductivity $\sigma (\omega)$ at $T=0$ for $D=0.8$ and $T_K=0.001$.}
\label{fig3}
\end{figure}

By expanding the previous expressions in the basis of the eigenstates 
$\left(\alpha ^{\dagger }\beta ^{\dagger }\gamma ^{\dagger }\right)$ of $H_{0}$,
we have computed the frequency dependence of $\chi ^{ab}({\bf {Q}},\omega )$ at the antiferromagnetic wavevector ${\bf {Q}}$ and $\sigma ^{ab}(\omega )$ at zero temperature. The two response functions show very different
frequency dependence. The frequency-scale at which the dynamical
spin susceptibility takes noticeable values is much smaller than for
the optical conductivity. This can be understood in the following
way. The bosonic $\gamma $-band is called to play a role only when
spin is concerned namely for the dynamical spin susceptibility. That 
feature comes from the fact that the spin is related to both
fermionic and bosonic operators while the c-current is simply
expressed within fermionic operators. Therefore, one can show that
the dynamical spin susceptibility involves transitions between all
three bands $\alpha $, $\beta $ and $\gamma $. The main
contribution for $\chi ^{ab}({\bf {Q}},\omega )$ is due to the
particle-hole pair excitations from the fermionic $\alpha $ to the
bosonic $\gamma $ band. Oppositely, the optical conductivity is associated with
transitions between fermionic bands only. As can be seen in Figure
\ref{fig3}, a gap appears in the frequency dependence of $\sigma (\omega )$
equal to the direct gap between the $\alpha $ and $\beta $ bands.
The latter result agrees with the predictions of the dynamical
mean-field theory in the limit of infinite dimensions \cite{rozenberg2}. 
The whole discussion clarifies the physical content of the novel bosonic 
mode brought by the supersymmetric approach. That
mode is related to the spin excitations. It introduces new features in
the dynamical spin susceptibility by comparison to the standard
slave-boson theories while it does not affect the optical conductivity.

\bigskip

\section{\protect\bigskip Conclusion}
Important progress has been made the last years in the understanding of the Kondo lattice model with the development of new functional integral approaches. They have enlightened as to the nature of the ground state and the existence of collective modes. They open up new prospects for the description of the critical phenomena associated to the quantum phase transition in Heavy-Fermion systems. A complete study of the quantum phase transition will probably requires the use of the Group Renormalization techniques for which the functional integral approaches presented here might constitute the framework.

\vspace{0.4in}
\centerline {\bf Acknowledgements}
\vspace{0.2in}
We would like to thank N.R. Bernhoeft, P. Coleman, P.A. Lee, G.G. Lonzarich, K. Miyake, M.J. Rozenberg, J. Spalek for many interesting and helpful discussions.
\bigskip 

$^*$ Also Part of the Centre National de la Recherche Scientifique (CNRS)

\vspace{0.4in}
\centerline {\bf APPENDIX}
\vspace{0.2in}

The expressions of the different bubbles appearing in the expression of the
boson propagators (cf. Eq.\ref{eq14}) are given here (with i=1, 2, m or ff)
\be
\overline{\varphi }_{i}({\bf{q}},i\omega_{\nu})=\varphi
_{i}({\bf{q}},i\omega_{\nu})+\varphi _{i}(-{\bf{q}},-i\omega_{\nu})
\ee
\begin{eqnarray*}
\varphi _{1}({\bf{q}},i\omega_{\nu}) &=&-\frac{1}{\beta} \sum_{k\sigma,i\omega_n}G_{cf_{0}}^{\sigma
}({\bf k+q},i\omega_{n}+i\omega_{\nu})G_{ff_{0}}^{\sigma }({\bf{k}},i\omega_{n}) \\
\varphi _{2}({\bf{q}},i\omega_{\nu}) &=&-\frac{1}{\beta} \sum_{k\sigma,i\omega_n}G_{cc_{0}}^{\sigma
}({\bf k+q},i\omega_{n}+i\omega_{\nu})G_{ff_{0}}^{\sigma }({\bf{k}},i\omega_{n}) \\
\varphi _{m}({\bf{q}},i\omega_{\nu}) &=&-\frac{1}{\beta} \sum_{k\sigma,i\omega_n}G_{cf_{0}}^{\sigma
}({\bf k+q},i\omega_{n}+i\omega_{\nu})G_{cf_{0}}^{\sigma }({\bf{k}},i\omega_{n})
\end{eqnarray*}
\begin{eqnarray*}
\varphi _{ff}^{\Vert }({\bf{q}},i\omega_{\nu}) &=&-\frac{1}{\beta } \sum_{k\sigma,i\omega_n}G_{ff_{0}}^{\sigma}({\bf k+q},i\omega_{n}+i\omega_{\nu})G_{ff_{0}}^{\sigma }({\bf{k}},i\omega_{n}) \\
\varphi _{cc}^{\Vert }({\bf{q}},i\omega_{\nu}) &=&-\frac{1}{\beta } \sum_{k\sigma,i\omega_n}G_{cc_{0}}^{\sigma }({\bf k+q},i\omega_{n}+i\omega_{\nu})G_{cc_{0}}^{\sigma }({\bf{k}},i\omega_{n})\\
\varphi _{fc}^{\Vert }({\bf{q}},i\omega_{\nu}) &=&-\frac{1}{\beta } \sum_{k\sigma,i\omega_n}G_{fc_{0}}^{\sigma }({\bf k+q},i\omega_{n}+i\omega_{\nu})G_{fc_{0}}^{\sigma }({\bf{k}},i\omega_{n})
\end{eqnarray*}
\begin{eqnarray*}
\varphi _{ff}^{\bot }({\bf{q}},i\omega_{\nu}) &=&-\frac{1}{\beta } \sum_{k\sigma,i\omega_n}G_{ff_{0}}^{\uparrow }({\bf k+q},i\omega_{n}+i\omega_{\nu})G_{ff_{0}}^{\downarrow }({\bf{k}},i\omega_{n}) \\
\varphi _{cc}^{\bot }({\bf{q}},i\omega_{\nu}) &=&-\frac{1}{\beta } \sum_{k\sigma,i\omega_n}G_{cc_{0}}^{\uparrow }({\bf k+q},i\omega_{n}+i\omega_{\nu})G_{cc_{0}}^{\downarrow }({\bf{k}},i\omega_{n}) \\
\varphi _{fc}^{\bot }({\bf{q}},i\omega_{\nu}) &=&-\frac{1}{\beta } \sum_{k\sigma,i\omega_n}G_{fc_{0}}^{\uparrow }({\bf k+q},i\omega_{n}+i\omega_{\nu})G_{fc_{0}}^{\downarrow }({\bf{k}},i\omega_{n})
\end{eqnarray*}
where $G_{cc_{0}}^{\sigma }({\bf{k}},i\omega_{n})$, $%
G_{ff_{0}}^{\sigma }({\bf{k}},i\omega_{n})$ and $G_{fc_{0}}^{\sigma }({\bf{k}},i\omega_{n})$ are the Green's
functions at the saddle-point level obtained by inversing the matrix $%
G_{0}^{\sigma }(\bf{k},\tau )$ defined in Equation (\ref{eq7}).

\vfill\eject


\bigskip

\begin{references}

\bibitem{lee} For a review see P.A. Lee, T.M. Rice, J.W. Serene, L.J. Sham, J.W. Wilkins, Comments Condens. Matter Phys. {\bf 12}, 99 (1986); D.M. Newns and N. Read, Adv.Phys. {\bf 36}, 799 (1987); P. Fulde, J. Keller, G. Zwicknagl, in Solid State Physics, edited by H. Ehrenreich and D. Turnbull (Academic, New York, 1988) {\bf 41}; N. Grewe and F. Steglich, in Handbook on the Physics and Chemistry of the Rare Earths (North Holland, Amsterdam, 1990), {\bf 14}

\bibitem{julian92} S.R. Julian, P.A.A. Teunissen, S.A.J. Wiegers, Phys. Rev. B {\bf 46}, 9821 (1992); S.R. Julian, F.S. Tautz, G.J. McMullan, G.G. Lonzarich, PhysicaB {\bf {199 \& 200}}, 63 (1994) 

\bibitem{doniach} S. Doniach, Physica B {\bf 91}, 231 (1977)

\bibitem{lohneysen} H.von L\"ohneysen, A. Shr\"oder, M. Sieck, T. Trappmann, 
Phys.Rev.Lett. {\bf 72}, 3262 (1994); H.von L\"ohneysen, Proceedings of the ITP Conference on Non-Fermi liquid behaviour in metals (Santa-Barbara June 1996) J.Phys.Cond.Matt. {\bf 8}, 9689 (1996)

\bibitem{julian96} S.R. Julian, C. Pfleiderer, F.M. Grosche, N.D. Mathur, G.J. McMullan, A.J. Diver, I.R. Walker, G.G. Lonzarich, Proceedings of the ITP Conference on Non-Fermi liquid behaviour in metals (Santa-Barbara June 1996) J.Phys.Cond.Matt. {\bf 8}, 9675 (1996) 

\bibitem{steglich} F. Steglich, B. Buschinger, P. Gegenwart, M. Lohmann, R. Helfrich; C. Langhammer, P. Hellmann, L. Donnevert, S. Thomas, A. Link, C. Geiber, M. Lang, G. Sparn, W. Assmus, Proceedings of the ITP Conference on Non-Fermi liquid behaviour in metals (Santa-Barbara June 1996) J.Phys.Cond.Matt. {\bf 8}, 9909 (1996) 

\bibitem{kambe} S. Kambe, S. Raymond, L.P. Regnault, J. Flouquet, P. Lejay, 
P. Haen, J.Phys.Soc.Jpn {\bf 65}, 3294 (1996)

\bibitem{cox} D.L. Cox Phys.Rev.Lett. {\bf 59}, 1240 (1987); D.L. Cox, M. Jarrell, Proceedings of the ITP Conference on Non-Fermi liquid behaviour in metals (Santa-Barbara June 1996) J.Phys.Cond.Matt. {\bf 8}, 9825 (1996) 

\bibitem{miranda} V. Dobrosavljevic, T.R. Kirkpatrick, G. Kotliar, Phys.Rev.Lett. {\bf 69}, 1113 (1992); E. Miranda, V. Dobrosavljevic, G. Kotliar Proceedings of the ITP Conference on Non-Fermi liquid behaviour in metals (Santa-Barbara June 1996) J.Phys.Cond.Matt. {\bf 8}, 9871 (1996) 

\bibitem{hertz} J.A. Hertz, Phys.Rev. B {\bf 14}, 1165 (1976)

\bibitem{millis93} A.J. Millis, Phys.Rev. B {\bf 48}, 7183 (1993)

\bibitem{moriya} T. Moriya, T. Takimoto, J. Phys. Soc. Jpn {\bf 64}, 960 (1995)

\bibitem{continentino} M.A. Continentino, Phys.Rev. B {\bf 47}, 11581 (1993)

\bibitem{rosch} A. Rosch, A. Schr\"oder, O. Stockert, H.v. L\"ohneysen, Phys. Rev. Lett. {\bf 79}, 159 (1997)

\bibitem{pepin} C. P\'epin and M. Lavagna, to be published in Phys.Rev. B

\bibitem{regnault} L.P. Regnault, W.A.C. Erkelens, J. Rossat-Mignod, P. Lejay, 
J. Flouquet, Phys. Rev. {\bf B 38}, 4481 (1988); S. Raymond, L.P. Regnault, S. Kambe, J.M. Mignod, P. Lejay, J. Flouquet, J. Low Temp.Phys. {\bf 109}, 205 (1997)

\bibitem{aeppli} G. Aeppli,
C. Broholm in Handbook on the Physics and Chemistry of Rare Earths  {\bf 19}, 123 (ed. by
Gschneidner et al, Elsevier 1994) and references within 

\bibitem{raymond} S. Raymond, L.P. Regnault, S. Kambe, J.M. Mignod, P. Lejay,
J. Flouquet, J.Low Temp.Phys. {\bf 109}, 205 (1997)

\bibitem{anderson} P.W. Anderson, Phys.Rev. B {\bf 124}, 41 (1961)

\bibitem{tsunetsugu} H. Tsunetsugu, M. Sigrist, K. Ueda,  Rev.Mod.Phys. {\bf 69},
809 (1997) and references within

\bibitem{nozieres} P. Nozi\`eres, Ann.Phys.Fr. {\bf 10}, 19 (1985) 

\bibitem{coleman} P. Coleman, Phys.Rev. B {\bf 29}, 3035 (1984)

\bibitem{read83} N. Read, D.N. Newns, J.Phys.C {\bf 16}, 3273 (1983)

\bibitem{millis87} A.J. Millis, P.A. Lee, Phys.Rev. B {\bf 35}, 3394 (1987)

\bibitem{auerbach86} A. Auerbach, K. Levin, Phys.Rev.Lett. {\bf 57}, 877 (1986)

\bibitem{lavagna87} M. Lavagna, A.J. Millis, P.A. Lee Phys.Rev.Lett. {\bf 58},
266 (1987)

\bibitem{read84} N. Read, D.M. Newns, S. Doniach, Phys.Rev.B {\bf 30}, 3841 (1984)
 
\bibitem{kroha} J. Kroha, P. Wolfle, T.A. Costi, Phys.Rev.Lett. {\bf 79}, 261 (1997)
 
\bibitem{houghton} A. Houghton, N. Read, H. Won, Phys.Rev. B {\bf 37}, 3782 (1988)

\bibitem{kotliar} G. Kotliar, A.E. Ruckenstein, Phys.Rev.Lett. {\bf 57}, 1362 (1986)

\bibitem{rice} T.M. Rice and K. Ueda, Phys.Rev.Lett. {\bf 55}, 995 (1985)

\bibitem{li} T. Li, P. Wolfle, P.J. Hirschfeld, Phys.Rev. B {\bf 40}, 6817 (1989)

\bibitem{lavagna90} M. Lavagna, Phys.Rev. B {\bf 41}, 142 (1990) 

\bibitem{yang} M.F. Yang, S.J. Sun, T.M. Hong, Phys.Rev. B {\bf 48}, 16123 (1993)

\bibitem{gulacsi} Zs. Gulacsi, R. Strack, D. Vollhardt, Phys.Rev. B {\bf 47}, 8594 (1993)

\bibitem{sun} S.J. Sun, M.F. Yang, T.M. Hong, Phys.Rev. B {\bf 48}, 16127 (1993)

\bibitem{jarrell} M.Jarrell, H. Akhlaghpour, T. Pruschke, Phys.Rev.Lett. {\bf 70}, 1670 (1993)

\bibitem{rozenberg} M.J. Rozenberg, Phys.Rev. B {\bf 52}, 7369 (1995)

\bibitem{doradzinski} R. Doradzi\'nski and J. Spalek, Phys.Rev. B {\bf 56}, R14239(1997); R. Doradzi\'nski and J. Spalek, Phys.Rev. B {\bf 58}, 1(1998)

\bibitem{moller} B. Moller and P. Wolfle, Phys.Rev. B {\bf 48}, 10320 (1993) 

\bibitem{lacroix} C. Lacroix and M.Cyrot, Phys.Rev.B {\bf 20}, 1969 (1979)
 
\bibitem{andrei} P. Coleman and N. Andrei, J. Phys.: Condens. Matt. {\bf 1}, 4057 (1989) 

\bibitem{iglesias} J.R.Iglesias, C. Lacroix, B. Coqblin, Phys.Rev. B {\bf 56}, 11820 (1997)

\bibitem{lavagna97} C. P\'epin and M. Lavagna, Z.Phys.B {\bf 103}, 259 (1997); C. P\'epin and M. Lavagna Cond-mat9709256

\bibitem{bernhoeft} N.R. Bernhoeft, G.G. Lonzarich, J.Phys.Cond.Mat. {\bf 7}, 7325 (1995)

\bibitem{kuramoto90} Y. Kuramoto, K. Miyake, J. Phys.Soc.Jpn {\bf 59}, 2831
(1990)

\bibitem{auerbach88} A. Auerbach, Ju H. Kim, K. Levin, M.R. Norman, Phys. Rev.
Lett. {\bf 60}, 623 (1988)

\bibitem{efetov} K.B.Efetov, Adv. Phys. {\bf 32}, 53 (1983)

\bibitem{gan} J.Gan, P.Coleman, N.Andrei, Phys.Rev.Lett. {\bf 68}, 3476 (1992) 

\bibitem{martin} R.M. Martin, Phys.Rev.Lett. {\bf 48}, 362 (1982)

\bibitem{rozenberg2} M.J. Rozenberg, G. Kotliar, H. Kajueter, Phys.Rev. B {\bf 54}, 8452 (1996) 

\end{references}
\end{document}